\documentclass[10pt]{article}
\usepackage[tbtags]{amsmath}
\usepackage{hyperref}
\usepackage{amssymb,amsthm}
\usepackage{amsfonts}
\usepackage{graphicx}
\graphicspath{{figs/}}
\usepackage[left=2.5cm,right=2.2cm,top=0.5cm,bottom=1.3cm,includeheadfoot]{geometry}
\usepackage{indentfirst}
\usepackage{color}
\usepackage{cite}
\usepackage{mathtools}
\usepackage{graphicx}
\usepackage{subcaption}
\usepackage{tikz-cd}
\usepackage{enumitem}

\definecolor{cites}{rgb}{0.65, 0.20, 0.46}
\definecolor{chapters}{rgb}{0.25, 0.30, 0.56}
\definecolor{header}{rgb}{0.0, 0.5, 0.64}

\hypersetup{
	colorlinks=true,
	linkcolor=header,
	citecolor=cites,
	urlcolor=chapters
}

\usepackage[normalem]{ulem}

\newcommand{\beq}{\begin{equation}}
\newcommand{\ee}{\end{equation}}

\numberwithin{equation}{section}

\renewcommand*{\thefootnote}{\fnsymbol{footnote}}

\begin{document}

\begin{center}
{\Large\bf
                How much chaos can be generated by gravitational collapse before reaching the Planck scale?
}
		\vskip 5mm
		{\large
			David Brizuela${}^{1,}$\footnote{Contact author: {\tt david.brizuela@ehu.eus}},
			Sara F. Uria${}^{2,}$\footnote{Contact author: {\tt sara.fernandezu@ehu.eus}},
			and Edward Wilson-Ewing${}^{3,}$\footnote{Contact author: {\tt edward.wilson-ewing@unb.ca}}
		}
		\vskip 3mm
		{\sl ${}^{1}$Department of Physics and EHU Quantum Center, University of the Basque Country,\\
			Barrio Sarriena s/n, 48940 Leioa, Spain}\\
			{\sl ${}^{2}$ Department of Applied Mathematics and EHU Quantum Center, University of the Basque Country,\\
	Plaza Ingeniero Torres Quevedo, 48013 Bilbao, Spain}\\
			{\sl ${}^{3}$ Department of Mathematics and Statistics, University of New Brunswick, \\
            Fredericton, NB,  E3B 5A3, Canada}\\
\end{center}

\setcounter{footnote}{0}
\renewcommand*{\thefootnote}{\arabic{footnote}}

\begin{abstract}
According to the Belinski-Khalatnikov-Lifshitz (BKL) conjecture the dynamics of general relativity near singularities is highly chaotic. However, since general relativity breaks down at singularities, it is generally expected that a more fundamental theory, such as quantum gravity, is needed to
describe the spacetime dynamics
in regions with large spacetime curvature.
In the absence of a widely accepted theory of quantum gravity, it remains unclear how exactly quantum effects may modify the classical evolution.
Taking a 
conservative point of view, 
in this paper we study the classical dynamics of a gravitational collapse, starting from a strong-field (though classical) scenario up to the Planck scale to quantify how much chaos is generated during the time
the Einstein equations can be trusted.
Specifically, we use
the Shannon entropy and Kullback-Leibler divergence, as well as Fourier analysis, to find that,
when the Planck scale is reached,
the chaotic features of the model remain relatively underdeveloped.
This suggests that, at least during the classical regime, chaos is not strong enough to erase all information about the initial state of the universe, or about a previous classical universe in the context of bouncing cosmologies. 
In addition, we also characterize the final invariant phase-space distribution for the chaotic Bianchi~IX dynamics in general relativity,
which provides a novel description of its invariant repeller. 
\end{abstract}

\section{Introduction}

The description of spacetime singularities remains one of the central open problems in
gravitational physics. As established by the singularity theorems \cite{Penrose:1965,Hawking:1970zqf}, in the context of general
relativity (GR), singularities arise generically under reasonable physical assumptions. However, these
theorems only guarantee geodesic incompleteness and provide little information about the detailed
dynamics of the spacetime as the singularity is approached. Understanding this asymptotic regime is of
great importance, both because it characterizes the ultimate predictions of GR and also because it concerns
the domain where modifications of GR due to quantum effects are typically expected to become relevant.

The Belinski-Khalatnikov-Lifshitz (BKL) conjecture \cite{Belinsky:1970ew}
provides a more detailed picture about the approach to spacelike singularities.
According to the conjecture, in the approach to a spacelike singularity the spacetime dynamics becomes asymptotically ultralocal; that is, time derivatives dominate over spatial derivatives, so that each typical spatial point evolves independently of its neighbors. Consequently, at a generic spatial point the Einstein field equations reduce to those of a homogeneous (but in general anisotropic) cosmology,
which can be classified according to the Bianchi models, with Bianchi~VIII and IX exhibiting the richest dynamics. Since this conjecture is based on distinguishing between space and time derivatives within a particular spacetime foliation, it is therefore not manifestly covariant; still, extensive analytical and numerical evidence strongly supports its validity as an asymptotic description of generic spacelike singularities \cite{Berger:1993ff, Berger:1998vxa, Berger:1998wr, Berger:2002st, Garfinkle:2003bb, Garfinkle:2004ww}. In this context, more recent studies found the presence of points where spatial gradients become temporarily significant and produce {\it spikes}, sharp features in the metric variables; however, these points lie on a two-dimensional surface within the three-dimensional spatial slice, and appear to be a set of measure zero \cite{Garfinkle:2020lhb, Heinzle:2012um}.

Further, according to the BKL conjecture, to leading order, ordinary matter fields (with the exception of a massless scalar field \cite{Belinski:1973zz}) become dynamically negligible compared with the gravitational degrees of freedom, so that the qualitative asymptotic behavior is essentially that of the vacuum Einstein equations for a Bianchi spacetime at each typical point: ``matter does not matter'' \cite{Ringstrom:2000mk}. However, this is strictly true only asymptotically and, for finite times, matter fields may modify certain properties of the BKL evolution \cite{Ali:2017qwa, Brizuela:2024ggl, Muzammil:2025nuv}.

The dynamics of Bianchi~VIII and Bianchi~IX models are qualitatively similar, so, for concreteness, here we will focus on the vacuum Bianchi~IX cosmological spacetime, also known as the Mixmaster universe \cite{Misner:1969ae, Misner:1969hg}. In the approach to the singularity,
the dynamics can be described as an oscillatory sequence of vacuum Bianchi~I solutions called Kasner epochs.
During each Kasner epoch the spatial curvature is negligible, and
the transitions between successive Kasner epochs occur when the spatial curvature becomes large for a short period of time.
In fact, the spatial curvature can be understood as a hard wall off which the system bounces almost instantaneously, and, consequently,
this oscillatory behavior is commonly referred to as cosmological billiards \cite{Damour:2002et}.
These bounces give rise to a highly intricate evolution that is characterized by sensitive dependence on initial conditions and a fractal, self-similar structure in phase space, and thus exhibits deterministic chaos \cite{Cornish:1996hx, Cornish:1996yg, Motter:2000bg, Heinzle:2009du}.

In summary, according to the BKL conjecture, the asymptotic approach to a spacelike singularity is typically ultralocal, vacuum-dominated, oscillatory, and chaotic. Nonetheless, it is important to emphasize that the BKL conjecture concerns the asymptotic limit towards the singularity, rather than the dynamics at any finite curvature scale. In classical GR there is no intrinsic fundamental length scale, and the neglect of spatial gradients and matter degrees of freedom removes any preferred scale from the equations, naturally leading to a scale-invariant dynamics.
This fact raises an important conceptual question: since the BKL regime is only attained asymptotically, to what extent are its characteristic chaotic properties actually realized before classical GR itself ceases to be applicable?

After all, it is widely expected that GR provides an accurate description of gravity only up to sufficiently high-curvature scales, beyond which new physical effects should become relevant. In particular, quantum effects are expected to become important at the Planck scale, but deviations from GR could arise at lower energies depending on the underlying theory. 
In general, such underlying theories are assumed to resolve the singularity in one way or another.
In cosmology, for example, some proposals replace the singularity by a
quantum tunneling from nothing event \cite{Vilenkin1982}, or by a
quantum state defined through a path integral over compact and regular
Euclidean four-geometries \cite{HartleHawking1983}, while still others
predict a nonsingular cosmological bounce \cite{Brandenberger:2016vhg}, including Einstein-Cartan
theory \cite{Poplawski2012}, loop quantum cosmology
\cite{AshtekarPawlowskiSingh2006}, matter-bounce scenarios
\cite{CaiBrandenbergerZhang2011}, and nonsingular ekpyrotic cosmologies
\cite{BuchbinderKhouryOvrut2007, Lehners:2008vx}.
Despite their conceptual differences, these proposals all share a common feature: they all only modify the classical dynamics after some finite cutoff
scale is reached. Consequently, irrespective of the ultimate fate of the singularity, the Einstein equations are expected to remain valid up to that particular scale.

This observation raises the question of how strongly the chaotic features of the BKL conjecture can affect the dynamics before the classical description breaks down. That is precisely the goal of this paper. To quantify this, we assume that (i) the BKL conjecture holds, and we focus on generic points that can each be described by a Mixmaster (vacuum Bianchi~IX) model, and (ii) the evolution is given by the Einstein equations. Then, the Shannon entropy \cite{ShannonWeaver:1949} and the Kullback-Leibler divergence \cite{KullbackLeibler:1951} can be used to quantify the distance at finite time of the system expressed in terms of a probability distribution from its invariant final distribution. In addition, the Fourier decomposition of the evolving state characterizes how information is transferred across scales.

Furthermore, this study will also shed light on the question of how strongly the BKL dynamics can enhance spacetime inhomogeneities in
the approach to a spacelike singularity. 
If the evolution of neighboring points decouples, and the dynamics at each point is chaotic, one may expect that nearby points (that have similar initial conditions) will eventually be driven to solutions far away from each other, and that this would produce very large inhomogeneities in the metric (with the inhomogeneities presumably diverging at the singularity).
However, if the chaotic evolution lasts only for a finite time (up to the cutoff scale where new physics arises), then perhaps the
amplitude of such inhomogeneities would remain bounded, at least up to that cutoff scale.

In fact, it has already been shown that there are typically only a few transitions between Kasner epochs in a collapse
from a macroscopic scenario until the time when the spatial volume of a Bianchi~IX universe becomes comparable to the Planck volume $\ell_{\rm Pl}^3$ \cite{Doroshkevich}.
This fact limits the amount of chaos that can be generated up to this point. However, within the framework of the BKL conjecture,
it is not obvious how to relate the spatial volume of a Bianchi~IX cosmology to the local physics at a single point,
and it seems that the curvature scale (not the spatial volume) is a more appropriate quantity to track in this context.
This is how we will proceed in order to estimate the duration in $e$-folds for a generic gravitational collapse
to follow the BKL behavior.

The outline of the paper is as follows. Since the dynamics at each (typical) point is assumed to be given by a
vacuum Bianchi~IX model, we start by briefly reviewing the Mixmaster dynamics in Sec.~\ref{sec-Mixmaster}. Then, in Sec.~\ref{sec-quantitative} 
we use
the Shannon entropy, the Kullback-Leibler divergence, and a Fourier analysis to 
quantify
how rapidly the 
chaotic features of the Mixmaster dynamics are developed during the
finite time interval between a high-curvature but classical initial scenario and the Planck scale. In Sec.~\ref{sec.invariant} we focus on the invariant final state,
reached once the chaos has fully emerged, in order to
characterize and summarize its statistical properties from a different perspective than is usually adopted. We end
the main body of the paper with a Discussion in Sec.~\ref{sec-disc}. Finally, in the appendices we present certain technical details of the study.
More precisely, in App.~\ref{sec:app.initial} we explain the choice of initial data for a sharply peaked distribution
in phase space (whose evolution is the one considered in the main body of the paper),
in App.~\ref{sec.Kretschmann} we provide the Kretschmann curvature scalar for Bianchi IX, while
in App.~\ref{sec_randomstate} we consider
an alternative initial uniform distribution over phase space 
to show the robustness of the results.

\section{Mixmaster dynamics}
\label{sec-Mixmaster}

As explained above, according to the BKL conjecture, for a typical point near a spacelike singularity, spatial gradients are negligible with respect to time derivatives, such that spatial points decouple, and in addition most common forms of matter become dynamically negligible. For this reason, the dynamics at a typical point is then given by a vacuum Bianchi model, most generally Bianchi~VIII or Bianchi~IX. These two spacetimes have similar dynamics, so here we will focus on the diagonal vacuum Bianchi~IX, also called the Mixmaster model.
This is a spatially homogeneous but anisotropic spacetime 
with topology $\mathbb{R}\times S^3$, and it is described by the metric
\begin{equation}
	\label{def_metric}
ds^2=-N(t)^2 dt^2+\gamma_{ij}\boldsymbol{\sigma}^i \boldsymbol\sigma^j,
\end{equation}
where $N(t)$ is the lapse function and $\boldsymbol\sigma^i$ ($i=1,2,3$) are the left-invariant one-forms on $SU(2)$, while
\begin{align}\label{def_gamma}
	\gamma_{ij}=\frac{r_0^2}{4}{\rm diag}\left(a_1(t)^2,a_2(t)^2,a_3(t)^2\right).
\end{align}
The scale factors $a_i(t)$ are dimensionless, while $r_0$ is the radius of the reference 3-sphere.

The vacuum
Einstein equations for the Bianchi~IX spacetime yield acceleration equations for each of the three scale factors,
\begin{align}\label{Einstein_equations}
	\begin{aligned}
		&\ddot{a}_1=
		-\frac{\dot{a}_1}{2}\left(\frac{\dot{a}_2}{a_2}+\frac{\dot{a}_3}{a_3}\right)
		+\frac{a_1}{2}\frac{\dot{a}_2}{a_2}\frac{\dot{a}_3}{a_3}
		+\dot{a}_1\frac{\dot{N}}{N}+
		\frac{a_1N^2}{
			2a^6r_0^2
		}\left(
			3(a_2^2-a_3^2)^2
			+2a_1^2(a_2^2+a_3^2)-5a_1^4
			\right),
		\\[8pt]
	&\ddot{a}_2=
	-\frac{\dot{a}_2}{2}\left(\frac{\dot{a}_1}{a_1}+\frac{\dot{a}_3}{a_3}\right)
	+\frac{a_2}{2}\frac{\dot{a}_1}{a_1}\frac{\dot{a}_3}{a_3}
	+\dot{a}_2\frac{\dot{N}}{N}+
	\frac{a_2N^2}{
		2a^6r_0^2
	}\left(
	3(a_1^2-a_3^2)^2
	+2a_2^2(a_1^2+a_3^2)-5a_2^4
	\right),
		\\[8pt]
	&\ddot{a}_3=
	-\frac{\dot{a}_3}{2}\left(\frac{\dot{a}_1}{a_1}+\frac{\dot{a}_2}{a_2}\right)
	+\frac{a_3}{2}\frac{\dot{a}_1}{a_1}\frac{\dot{a}_2}{a_2}
	+\dot{a}_3\frac{\dot{N}}{N}
	+\frac{a_3N^2}{
		2a^6r_0^2
	}\left(
	3(a_1^2-a_2^2)^2
	+2a_3^2(a_1^2+a_2^2)-5a_3^4
	\right),
	\end{aligned}
\end{align}
together with the constraint
\begin{align}\label{constraint}
	-2\left(H_1H_2
		+H_1H_3+
		H_2H_3
		\right)
		+\frac{ a_1^4 +(a_2^2-a_3^2)^2 -2a_1^2(a_2^2+a_3^2) }{ a^6r_0^2 }=0,
\end{align}
where a dot denotes a derivative with respect to the generic time $t$, $a = (a_1 a_2 a_3)^{1/3}$ is the (geometric) mean scale factor,
$H_i= \dot{a}_i/(a_i N)$, with $i=1,2,3$, are the directional Hubble factors, while $H = \dot{a}/(aN)$ is the mean Hubble rate.
In the constraint \eqref{constraint}, the first terms, which are quadratic in the velocities $H_i$, can be regarded as the kinetic part,
while the remainder is due 
to the three-dimensional Ricci scalar ${}^{(3)}\!R $ and can be interpreted as
a potential. 
Such spatial curvature is also responsible for the (force-like) terms proportional to $1/r_0^2$ in the acceleration equations \eqref{Einstein_equations}.

\begin{figure}
	\centering
	\includegraphics[width=0.8\linewidth]{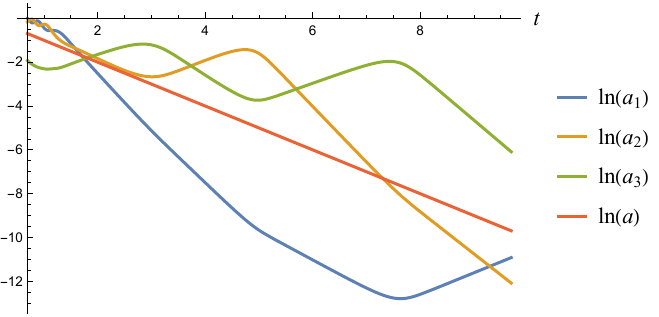}
	\caption{Evolution of the scale factors $a_i$ together with the average scale factor $a$ in logarithmic scale with respect to $t=-\ln a$ as the singularity (located at  $t \to +\infty$) is approached.}
	\label{fig:evolution_scale_factors}
\end{figure}

Evolving towards the cosmological singularity at $a(t)=0$, the mean scale factor decreases monotonically, but the directional scale factors have richer dynamics: two of the directional scale factors contract, while the third one expands, with a periodic exchange of roles. This oscillatory behavior is shown in Fig.~\ref{fig:evolution_scale_factors}.

Away from the turning points in the oscillations, the potential due to the spatial curvature
is negligible and the solutions for the scale factors have a simple form in cosmic time ${\tilde t}$ (with $N=1$),
namely $a_i(t) \sim \tilde{t}^{k_i}$, where the Kasner exponents $k_i$ satisfy $\sum k_i = \sum k_i^2 = 1$.
Note that two $k_i$ are necessarily positive, while the other one must be negative, so in the approach to the singularity (in this time gauge as $\tilde t \to 0$) two directions are contracting and the other one is expanding.
These solutions correspond to the vacuum Bianchi~I (or Kasner) spacetime, and provide a good approximation to the full dynamics during the time intervals in which the three directional scale factors evolve monotonically. For this reason, each such interval is called a Kasner epoch, and the oscillation between one Kasner epoch and another, when the potential is no longer negligible, is called a Kasner transition.

There are simple transition rules relating the Kasner exponents of subsequent Kasner epochs \cite{Belinsky:1970ew, Belinsky:1982pk}, and these have been shown to generate chaos \cite{Barrow:1981, Barrow:1981sx, Barrow_Chernoff:1983, Cornish:1996hx, Cornish:1996yg, Motter:2000bg, Imponente:2001fy}.  Further, it is also possible to numerically solve the exact Bianchi~IX equations to demonstrate that they are chaotic \cite{Cornish:1996hx, Cornish:1996yg, Motter:2000bg}. At the same time, this dynamics gradually erases memory of the initial state, a property known as mixing \cite{Belinski:2017fas}. Although chaos and mixing are distinct concepts, the mixing provides a physically intuitive measure of chaos by quantifying the loss of correlations during the evolution.

\subsection{Compactification of the phase space}
\label{sec_compactification}

In a chaotic dynamical system, it is sometimes the case that not all of the phase-space coordinates behave chaotically (for example, in the Bianchi~IX spacetime the dynamics of the mean scale factor $a(t)$ is not chaotic), and, in such case, it can be very helpful to isolate the key variables
that encode the essential chaotic dynamics. In addition, it is also convenient to compactify these coordinates so that all possible initial conditions (and all orbits) are bounded. Another important advantage of such a compactification is that it makes it easier to analyze how an initially localized distribution spreads over the accessible phase space and progressively loses memory of its initial state.

The phase space for the Bianchi~IX spacetime is six-dimensional and coordinatized by the directional scale factors $(a_1, a_2, a_3)$ and their velocities $(\dot{a}_1,\dot{a}_2,\dot{a}_3)$, which implies three degrees of freedom. However, the constraint \eqref{constraint} is first-class,
so one of the three degrees of freedom is gauge, as it is related to the freedom of the choice of the time variable.
In the following, since in the approach to the singularity (towards the high-curvature regime) $a(t)$ decreases monotonically,
we select the time variable $t=-\ln a$ to fix this gauge, so the singularity $(a\to 0)$ is located at $t\to\infty$.
The velocity $\dot{a}$ is then fixed by the constraint \eqref{constraint}, and the corresponding lapse reads $N=-1/H$.
This gauge choice gives a simple geometric interpretation for $t$: it is the number of $e$-folds of contraction of the mean scale factor.

Therefore, the physical phase space is four-dimensional and can be coordinatized in terms of two variables related to the directional scale factors $a_i$, and two other variables that capture their velocities $\dot{a}_i$. As explained above, it is convenient to use compact coordinates when possible,
so we can describe the four remaining variables in terms of two two-dimensional vectors in polar coordinates. Specifically, the moduli $\beta$ and $P$ are defined as follows,
\begin{align}\label{def_beta}
   \beta^2 &:=\frac{1}{12}
    \ln^2\left(
    \frac{a_1}{a_2}
    \right)+\frac{1}{4}\ln^2\left(
    \frac{a}{a_3}
    \right),\\[8pt]
    \label{def_P}
    P^2 &:=24r_0^2 a^6\left[
(H_1-H)^2+(H_2-H)^2+(H_3-H)^2
\right],
\end{align}
while the two angles $\varphi$ and  $\theta$ are given by 
\begin{align}
	\label{def_phi}
	\cos\varphi:=\frac{1}{2\beta}\ln\left(
\frac{a}{a_3}
	\right),\quad & \quad \sin\varphi=\frac{1}{\sqrt{12} \beta}\ln\left(
	\frac{a_1}{a_2}
	\right),
		\\[8pt]
	\label{def_theta}
	\cos\theta:=\frac{6 r_0 a^3}{P}\left(
	H-H_3
	\right),\quad & \quad \sin\theta=\frac{\sqrt{12} r_0 a^3}{P}(H_1-H_2).
\end{align}
For a complete derivation of these variables
we refer the reader to Apps.~\ref{sec:app.Misner} and \ref{sec:app.angles}.
As it is clear from these definitions, $\beta$ and $P$ measure the magnitude of the anisotropy of the directional scale factors and of the directional Hubble rates, respectively, while the angles $\varphi$ and $\theta$ capture how the anisotropies (again of the directional scale factors and of the directional Hubble rates, respectively) are distributed among the three directions.
All of these variables are dimensionless.

It has been shown that it is the angles $(\theta,\varphi)$ that are the relevant phase-space coordinates that encode the chaotic character of the model \cite{Cornish:1996hx,Cornish:1996yg,Barrow_Chernoff:1983}. For this reason, we will use this compact section of the phase space to describe the dynamics of the system, and to analyze and quantify its chaotic features.

\subsection{Validity range of the BKL dynamics}
\label{sec-range}

As discussed above, the BKL behavior only appears as a spacelike singularity is approached, but there is no a priori guarantee that this will occur at a certain scale: there is no ``BKL scale'' beyond which the BKL dynamics can generically be expected to hold.
Here we will be conservative in our estimate of how long this dynamics can last before reaching the Planck scale, at least for the types of curvature singularities that are of the greatest physical interest: black hole singularities and the initial cosmological singularity for our universe.

Since proposals for new gravitational physics (including quantum-gravity theories) typically start to depart from general relativity at a certain curvature scale, we will focus on a collapse from
an initial state corresponding to a classical
macroscopic scale, but with an already high curvature so the system is reasonably close to the singularity, to the final state at the Planck scale. (Note that some modified gravity theories would predict departures from general relativity before reaching the Planck scale, so this is a conservative choice for the maximum duration of the BKL scenario.)
Also, note that, since the BKL dynamics is vacuum-dominated, curvature invariants built from the Ricci tensor vanish and nontrivial curvature invariants
are entirely determined by the Weyl tensor. Thus, a convenient way to characterize the curvature during the collapse is through the Kretschmann scalar,
defined as $K = R_{abcd} R^{abcd}$ in terms of the Riemann tensor $R_{abcd}$.

To translate the difference in curvature scales to a time interval, keeping in mind that $t = - \ln a$, it is necessary to relate the Kretschmann scalar $K$ in the Bianchi~IX model (since we assume BKL dynamics holds for the entire duration being considered) to the number of $e$-folds in the mean scale factor. As the singularity is approached, $K$ scales as\footnote{See App.~\ref{sec.Kretschmann} for the complete form of the Kretschmann scalar
for the Bianchi IX spacetime. In fact, there is also a contribution $P^4$ to the numerator, but, while $P$ does evolve during the dynamics, it does not vary nearly as much as $a$, and, as a result, the change in $K$ is most strongly driven by changes in $a$, which is what we focus on here.}
\begin{align}\label{def_Kretsch}
	K \sim \frac{1}{(a^3 r_0)^4}.
\end{align}

Given how $K$ scales with respect to $a$, the number of $e$-folds experienced by the system when undergoing a contraction from an initial configuration with $K(t_0)$ to a final state with $K(t_f)$ is
\begin{align}\label{def_e_folds}
	N_{e\text{-folds}}:=\ln\left(
	\frac{a(t_0)}{a(t_f)}
	\right)=\frac{1}{12}\ln \left(
	\frac{K(t_f)}{K(t_0)}
	\right).
    \\[-10pt]
    \nonumber
\end{align}
As discussed above, we assume that the final curvature scale (that is, the curvature scale of new physics) is the Planck scale, so $K(t_f)=K_{\rm Pl}\sim {\ell_{\rm Pl}^{-4}}$, where $\ell_{\rm Pl}$ is the Planck length. 
We also note that the amount of $e$-folds $N_{e\text{-folds}}$ is independent of the radius of the 3-sphere $r_0$.
This parameter is not relevant dynamically, since, as commented above, the dynamics is scale invariant.
In fact, it is easy to see that $r_0$ can be absorbed into the lapse, so that it completely disappears
from the equations of motion \eqref{Einstein_equations}--\eqref{constraint}.
This parameter $r_0$ is only relevant when explicitly constructing the metric and its corresponding
curvature tensors and scalars.

The choice for the initial state is less obvious. To be conservative, we will consider the most extreme regions experimentally probed so far, where we have evidence that the BKL behavior is not yet present, both for black holes and in cosmology.

For black holes, gravitational wave observations of binary black hole mergers show physics where the black holes clearly interact, so the dynamics is not ultralocal as it would be for BKL behavior; these observations probe physics down to close to the horizons of the black holes. Many of the observations by the LIGO, Virgo, and KAGRA collaborations involve black holes with a mass on the order of a few solar masses, and at the horizon of a black hole of mass $M$
\beq
K \sim \left(\frac{M_{\rm Pl}}{M}\right)^{\!4}\ell_{\rm Pl}^{-4},
\ee
where $M_{\rm Pl}$ is the Planck mass.
Taking $M$ to be within an order of magnitude of a solar mass, we get
\begin{align}\label{Kretschammn_initial}
	K(t_0)\sim \left(\frac{M_{\rm Pl}}{M_{\rm sun}}\right)^{\!4}\ell_{\rm Pl}^{-4}\sim 10^{-152}\ell_{\rm Pl}^{-4},
\end{align}
for the initial estimate. Hence, in this case, following \eqref{def_e_folds}, the maximal number of $e$-folds possible before reaching the Planck scale is
\begin{align}
N_{e\text{-folds}}\approx \frac{1}{12}\ln\left(10^{152}\right) \sim 30.
\end{align}
If one were to take the black hole to be supermassive, for example, the one at the center of the Milky Way,
then $M \sim 10^6 M_{\rm sun}$, and this would give only an additional $\sim 5$ $e$-folds of contraction.

Note that this estimate provides a very conservative upper bound for the number of $e$-folds of contraction, as it assumes that the BKL dynamics begins immediately just inside the black hole horizon, and this is not realistic. A more accurate approximation would have the BKL regime start at larger curvatures, and then there would be fewer $e$-folds of contraction before reaching the Planck scale. 
Nonetheless, as shall be shown below, even this conservative upper bound is already sufficient to limit the amount of chaos that can occur before reaching the Planck scale.

A similar estimate can be obtained for the cosmological case.  In cosmology, the predictions of big bang nucleosynthesis have been confirmed.
During that epoch, the cosmological evolution is driven by a radiation fluid, and is not vacuum-dominated as it would be under BKL behavior.
Furthermore, the temperature (or energy) scales at which big bang nucleosynthesis occurs is from $T \sim 1$ MeV to $T \sim 10$ keV, so we assume that the smallest energy scale at which BKL behavior could occur in this context is at $T \sim 10$ MeV, which is about 21 orders of magnitude below the Planck temperature $T_{\rm Pl} \sim 10^{19}$ GeV. Since for a radiation-dominated universe $K = 12 (8 \pi G \rho / 3)^2$, where the energy density of the radiation fluid is $\rho \sim T^4 / \hbar^3$ with a prefactor of order 1 that depends on the number of species of particles, it follows that $K(t_0) \sim 10^{-165} \ell_{\rm Pl}^{-4}$.
If the BKL behavior begins just before big bang nucleosynthesis (again, this is not realistic, and is only used to obtain a conservative upper bound for the maximum duration of the BKL dynamics before reaching the Planck scale) then
\begin{align}
N_{e\text{-folds}}\approx \frac{1}{12}\ln\left(10^{165}\right)\sim 30.
\end{align}

Interestingly, the (conservative) upper bounds from both black holes and cosmology give a very similar result. Therefore, following these estimates, here we will assume there are at most 30 $e$-folds of contraction from the onset of the BKL behavior to the Planck scale. This should be understood as a conservative upper bound for systems that are cosmologically or astrophysically relevant (there may be other less observationally relevant solutions to general relativity for which there are more $e$-folds of contraction).

\section{Quantitative study of mixing at finite times}
\label{sec-quantitative}

The Bianchi~IX Mixmaster dynamics is commonly studied by following a trajectory in configuration or phase space, but here we will instead follow the evolution of a distribution in phase space.
A phase-space distribution (rather than a phase-space point) captures more directly the physics that we want to study for two main reasons. First, the initial conditions for a physical system are never known exactly, but rather contain some small uncertainty. For a system that is not chaotic, this uncertainty can typically be ignored since the trajectories of nearby solutions do not significantly diverge and it is sufficient to calculate the system's evolution for a particular trajectory, and allow for a small uncertainty around that trajectory. On the other hand, for a chaotic system, trajectories that are initially nearby will significantly diverge from one another, and then it becomes necessary to study how the distribution of possible initial conditions (given the initial uncertainty) evolves dynamically.

Second, given the underlying motivation of the BKL conjecture, we are also interested in how the geometry at nearby spacetime points changes; in particular, given an initial spatial hypersurface, if there is a certain region $R$ with small inhomogeneities, how do these inhomogeneities grow during a BKL regime? This question can be addressed by considering a phase-space distribution constructed from the values of the phase-space variables at different spatial points within $R$. If the inhomogeneities in $R$ are initially small, the corresponding distribution will be sharply peaked. As the system evolves, the initially nearby configurations may follow increasingly different trajectories in phase space, causing the distribution to spread. The growth of this spread therefore provides a measure of the growth of spatial inhomogeneities in $R$.

For these reasons, in the following we will study the Mixmaster dynamics of distributions in phase space, starting with an initially peaked state and then computing how it spreads through the phase space with time and eventually tends to an invariant final state---any initial distribution on the phase space asymptotically approaches this invariant statistical measure \cite{Barrow_Chernoff:1983,Cornish:1996hx,Cornish:1996yg}. Furthermore, in order to show the robustness of the
results, in App.~\ref{sec_randomstate} we take an initial uniform distribution over the phase space, and we find that eventually it also tends to the same invariant final state.

In this section, our goal is to quantitatively characterize the approach to the final invariant state by explicitly computing a set of markers that encode how mixed the state
is at a given finite time, and through this, how far it remains from the final invariant measure. In particular, as explained above, we will be interested in answering this question at the time the system reaches the Planck scale (after $\sim30$ $e$-folds of evolution, as explained in Sec.~\ref{sec-range}), at which point general relativity is typically expected to break down.

The remainder of the section is organized as follows. In Sec.~\ref{sec_initiallypeaked} we explain how initial conditions are chosen, and describe the evolution of these states in Sec.~\ref{sec_phase_space_evolution}; in particular we present several plots of the reduced phase space $(\theta,\varphi)$ at different evolution times and show how the distribution on phase space tends to the final invariant state. Then, in Sec.~\ref{sec_Basin} we define 3 basins and study how the evolution changes the basin the different initial conditions lie in. At the final invariant state the boundaries between the basins are known to be fractal and demonstrate the presence of chaos, but we find that the amount of mixing remains very limited when the Planck scale is reached.
We make this observation more quantitative in the subsequent sections:
in Sec.~\ref{sec_shannon} by computing the Shannon entropy of the distribution, that quantifies the degree of delocalization of the distribution and consequently the extent to which information about the initial state has been erased;
in Sec.~\ref{sec_KLdivergence} by calculating the Kullback-Leibler divergence to measure the distance of the state to the final invariant distribution;
and in Sec.~\ref{sec_Fourier} by studying the phase-space distribution in Fourier space to understand 
the distribution's degree of structure and the flow of information across scales at different times.

As explained in the latter part of the outline, this analysis will characterize (in addition to intermediate states) also the final invariant measure of the Bianchi~IX dynamics in a variety of ways, many of them new, and therefore in Sec.~\ref{sec.invariant} we will summarize and give further details about the final state's statistical properties in terms of these characterizations.

\subsection{Initial distribution}
\label{sec_initiallypeaked}

We will now specify the initial conditions for the distribution in phase space in terms of the compact variables $\theta$ and $\varphi$ introduced in Sec.~\ref{sec_compactification}.

We are interested in a distribution that is initially sharply peaked in terms of $(\theta,\varphi)$, so we discretize the phase space into a $30\times 30$ equally distributed cells 
in the $\theta-\varphi$ plane.
Then, we randomly choose one cell and populate it with $N=300^2$ points, randomly distributed within that particular cell.

Each of these points provide an initial value for $(\theta,\varphi)$, so there remain three additional initial conditions to be fixed: in addition to the modulus of the anisotropies $\beta$ and their velocity $P$, defined in \eqref{def_beta} and \eqref{def_P}, the value of the initial time $t_0$ must also be selected, since the energy constraint \eqref{constraint} depends on time
(note that the gauge choice $t=-\ln a$ implies that selecting $t_0$ is equivalent to selecting $a(t_0)$). In the following, we summarize how we define the initial conditions for these three variables; a more detailed description is given in App.~\ref{sec:app.initial}.

We will set conditions such that at the initial time the system is in a Kasner epoch.
For such a purpose, first, we select $\beta(t_0)$ under the condition that, for the given $(\theta(t_0), \varphi(t_0))$,
the potential term in the energy constraint \eqref{constraint} exactly vanishes.
Then, we need to take into account that the evolution of Bianchi~IX generically presents a recollapse at large
volumes \cite{Lin:1989tv} and that the characteristic BKL dynamics, with alternating Kasner epochs,
only arises sufficiently far from the recollapse.
So let us define the dimensionless ratio $\rho:=-12r_0 a H$, which increases
monotonically from the recollapse (where $\rho\to 0$) to the eventual singularity (where $\rho\to+\infty$).
In addition, whenever the potential term vanishes, as it is the case for our initial data, $\rho=P/a^2$.
Therefore, to ensure that the initial state lies sufficiently far from both the recollapse and the singularity,
we choose $a(t_0)$ and $P(t_0)$ such that $\rho(t_0)$ takes an appropriate intermediate value.
Following Refs.~\cite{Barrow_Chernoff:1983,Cornish:1996hx,Cornish:1996yg}, we select the initial condition
$a_1(t_0)=1$ for the first scale factor, which,
together with the chosen values for $\beta(t_0)$ and $\varphi(t_0)$, completely fixes $a(t_0)$ by the relation
\beq
a=a_1\,e^{-\beta(\sqrt{3}\sin\varphi+\cos\varphi)},
\ee
providing typical values of the order $a(t_0) \sim 10^{-1}$.
Concerning $\rho(t_0)$, based on an extensive numerical study, a reasonable range is approximately
$\rho(t_0) \in [4~000, 10~000]$, which, given the choice for $a(t_0)$, implies
$P(t_0) \in [40, 100]$. We choose $P(t_0)$ randomly within this range, and use the same value for
all of the points in the distribution.

\subsection{Evolution of the phase-space distribution}
\label{sec_phase_space_evolution}

\begin{figure}
	\centering
	
	\begin{subfigure}{0.48\textwidth}
		\centering
		\includegraphics[width=0.88\textwidth]{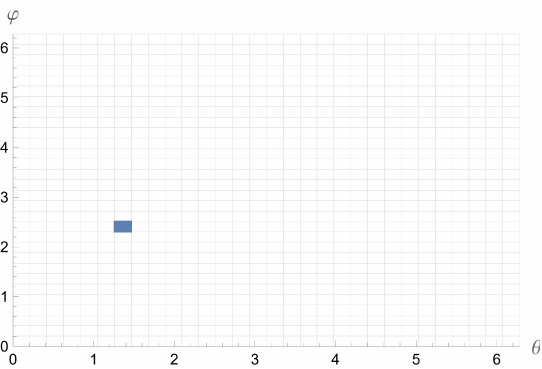}
		\caption{Initial state.}
        \label{fig:0_efolds}
	\end{subfigure}
	\hfill
	\begin{subfigure}{0.48\textwidth}
		\centering
		\includegraphics[width=0.88\textwidth]{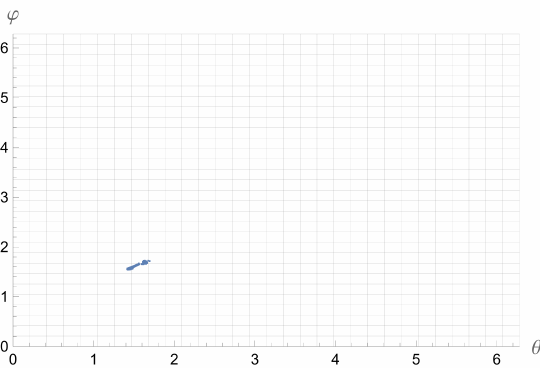}
		\caption{$1$ $e$-fold.}
        \label{fig:1_efolds}
	\end{subfigure}
		
	\begin{subfigure}{0.48\textwidth}
		\centering
		\includegraphics[width=0.88\textwidth]{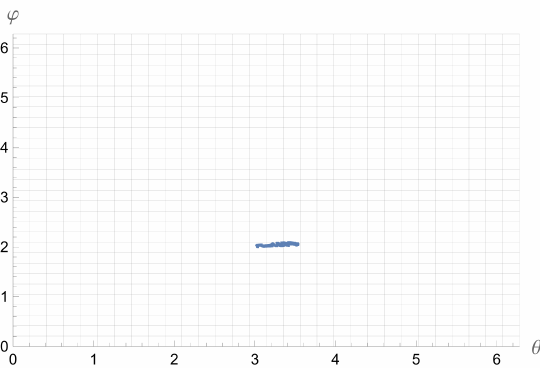}
		\caption{$2$ $e$-folds.}
        \label{fig:2_efolds}
	\end{subfigure}
	\hfill
	\begin{subfigure}{0.48\textwidth}
		\centering
		\includegraphics[width=0.88\textwidth]{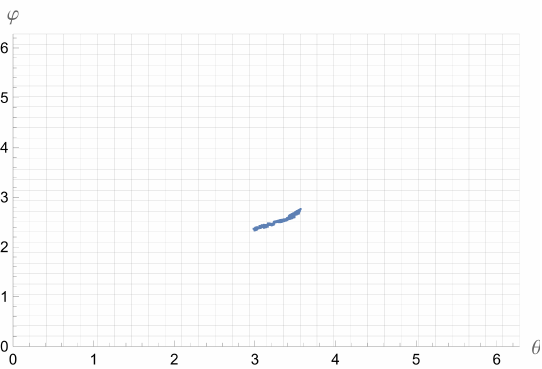}
		\caption{$4$ $e$-folds.}
        \label{fig:4_efolds}
	\end{subfigure}
	
	\begin{subfigure}{0.48\textwidth}
		\centering
		\includegraphics[width=0.88\textwidth]{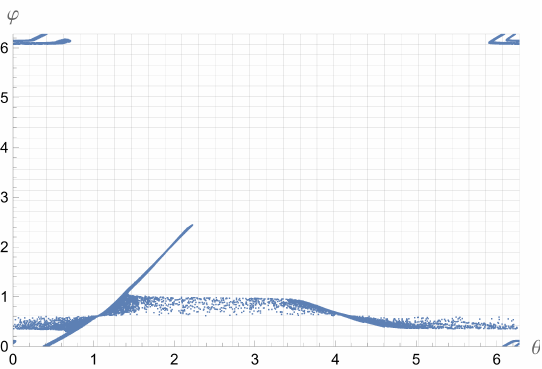}
		\caption{$30$ $e$-folds (Planck scale).}
        \label{fig:planck_limit}
	\end{subfigure}
	\hfill
	\begin{subfigure}{0.48\textwidth}
		\centering
		\includegraphics[width=0.88\textwidth]{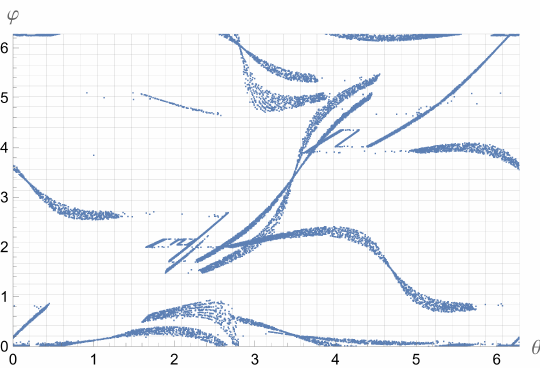}
		\caption{$200$ $e$-folds.}
       \label{fig:200_efolds}
	\end{subfigure}

    \begin{subfigure}{0.48\textwidth}
		\centering
		\includegraphics[width=0.88\textwidth]{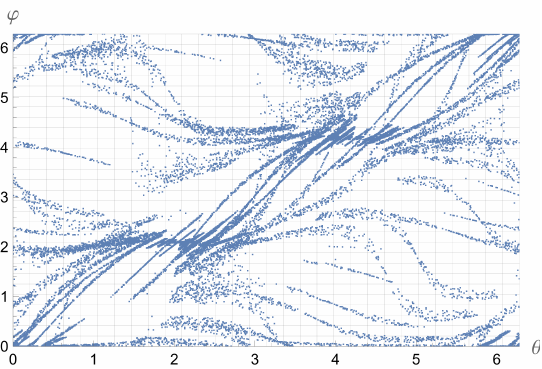}
		\caption{$1000$ $e$-folds.}
        \label{fig:1000_efolds}
	\end{subfigure}
	\hfill
	\begin{subfigure}{0.48\textwidth}
		\centering
		\includegraphics[width=0.88\textwidth]{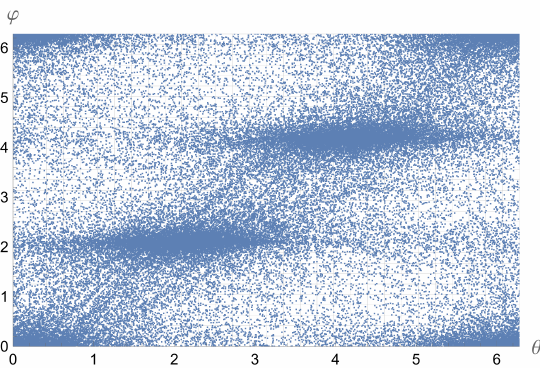}
		\caption{$60\,000$ $e$-folds (final invariant state).}
       \label{fig:late_evolution}
	\end{subfigure}

	\caption{Evolution of the distribution in $(\theta,\varphi)$ space at different stages of the dynamics. Each point corresponds
    to a different set of initial data.
    At the initial time (plot \ref{fig:0_efolds}) all points are inside a given cell, representing
    a sharply peaked distribution, which spreads as the spacetime collapses over different $e$-folds. Nonetheless, when the Planck scale
    is reached (plot \ref{fig:planck_limit}), the distribution still strongly differs from 
    the final invariant state (plot \ref{fig:late_evolution}).
}
	\label{fig:early_evolution}
\end{figure}
\begin{figure}
	\centering
	\includegraphics[width=0.65\linewidth]{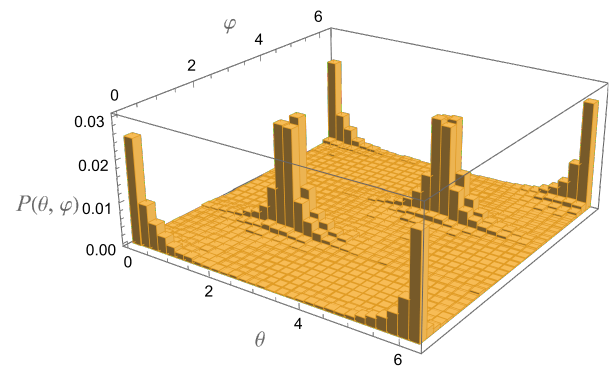}
	\caption{Normalized histogram representing the probability distribution of the invariant state $P(\theta,\varphi)$,
    defined as the ratio between the amount of points in a given cell and the total number of points $N$.}
	\label{fig:histogram}
\end{figure}
\begin{figure}[p]
	\centering
	\begin{subfigure}{0.48\textwidth}
		\centering
		\includegraphics[width=\textwidth]{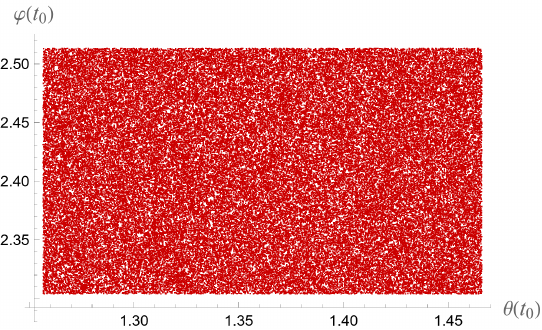}
		\caption{Initial state.}\label{fig:basins1}
	\end{subfigure}
	\hfill
	\begin{subfigure}{0.48\textwidth}
		\centering
		\includegraphics[width=\textwidth]{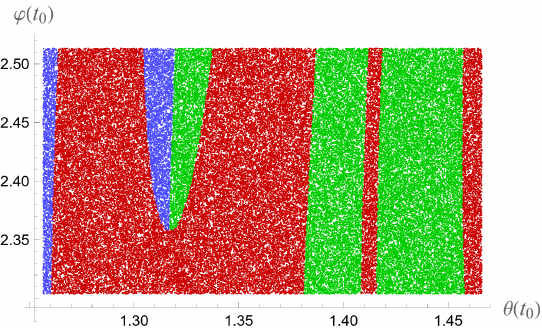}
		\caption{$30$ $e$-folds (Planck scale).}\label{fig:basins2}
	\end{subfigure}
    \vspace{0.4cm}
    
	\begin{subfigure}{0.48\textwidth}
		\centering
		\includegraphics[width=\textwidth]{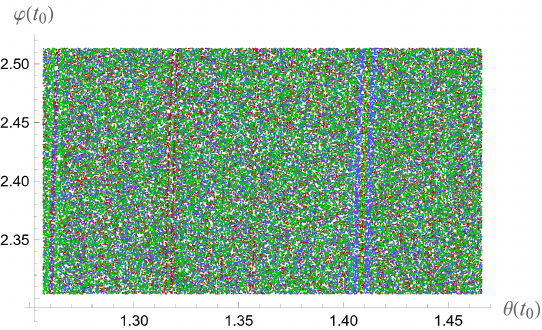}
		\caption{$60~000$ $e$-folds (final invariant state).}\label{fig:basins3}
	\end{subfigure}
	\caption{The basin boundaries for the initial distribution at different times. The colors red, blue, and green correspond to the three different basins defined in the text.
    In all the plots $(\theta(t_0),\varphi(t_0))$ correspond to the initial values of those variables.}
	\label{fig:basins}
\end{figure}

The phase-space distribution, with initial conditions chosen to be sharply peaked in phase space as described in Sec.~\ref{sec_initiallypeaked}, is then evolved numerically, point by point.
After a few $e$-folds of evolution (recall that $t = -\ln a$), the distribution begins to spread, but only mildly, with much of the initial structure still remaining as can be seen in Figs.~\ref{fig:0_efolds}--\ref{fig:4_efolds}. For larger $t$, the spreading continues to increase, but even
when the Planck scale is reached at $N_{e\text{-folds}}\sim 30$, the distribution remains far from being fully spread or mixed, as can be seen in Fig.~\ref{fig:planck_limit}.

To better assess the degree of mixing, the system can be evolved still further, as shown in Figs.~\ref{fig:200_efolds}--\ref{fig:late_evolution}.
After approximately 60~000 $e$-folds of contraction,
the distribution settles down to an invariant configuration,
at least within our numerical resolution. This final distribution is shown in Fig.~\ref{fig:late_evolution}. The final invariant distribution is not uniform: as is clearly shown in the histogram in Fig.~\ref{fig:histogram}, there are clusters in certain regions of phase space ($\theta,\varphi=0, 2\pi/3, 4\pi/3, 2\pi$), due to an underlying repelling structure that governs the dynamics and prevents uniform exploration of phase space (more details on this repeller can be found in Ref.~\cite{Cornish:1996hx}).

Importantly, the same late-time behavior is observed for different choices of initial cells and initial values of $P$ in the commented range $P(t_0)\in [40,100]$. The only differences appear in the initial transient, since at early $e$-folds the trajectories are in general quite different. However, after this transient regime, the distribution always converges to the same invariant state, regardless of the initial conditions. Moreover, in all cases the system still remains far from this final invariant state when it reaches the Planck scale.

\subsection{Basin boundaries}
\label{sec_Basin}

A coordinate-independent measure of chaos is given by studying the basin boundaries of the system. To do this, the phase space must be split into different basins, by determining which initial conditions end in which basin. Then, if the boundaries between the basins are fractal, the system is chaotic.
Note that, depending on the system, the basins can in some cases be states that are asymptotically approached, or alternatively can be the state of the system at some instant of time.
For our purposes, it is more natural to define the basins at an instant of time, which also makes it possible to see how and when the chaos appears.

For the Mixmaster dynamics, there are many different possible choices for basins, but for simplicity here we will define three basins,
depending on which of the three scale factors is expanding (recall that in a Kasner epoch, there is always one scale factor that is expanding, while the other two are contracting, as seen in Fig.~\ref{fig:evolution_scale_factors}). Then, at each time, it is possible to assign any point to one of the basins.

The results are shown in Fig.~\ref{fig:basins},
where in all the plots $(\theta(t_0),\varphi(t_0))$ correspond to the initial values of those variables, and the color of each point
represents the basin it corresponds to at different times. As can be seen, at the initial time (Fig.~\ref{fig:basins1}) all points are in the same basin.
When reaching the Planck scale after 30 $e$-folds (Fig.~\ref{fig:basins2}) there is some mixing, but the boundaries between the different basins are smooth curves (not fractals). This is because the trajectories have undergone only a few Kasner transitions at this stage, whereas
the development of chaotic properties typically requires of the order of tens of transitions. As commented above, chaos is fully developed once the invariant final state is reached,
after approximately 60~000 $e$-folds of contraction (Fig.~\ref{fig:basins3}). At this point
the boundaries have become fractal and, in fact, there is such a high degree of mixing that it is difficult to locate boundaries in the first place.

This series of plots clearly shows that it takes some time for the chaos to show itself, and the relatively short amount of time needed to reach the Planck scale does not seem to be sufficient by far. A more quantitative exploration of this observation is carried out in the next sections.

\subsection{Shannon entropy}
\label{sec_shannon}

Shannon entropy provides a quantitative description of the spread of the phase-space distribution. In particular,
given an initially sharply peaked distribution, the Shannon entropy can quantify how much information about the system’s initial state is lost. More precisely, Shannon entropy is defined as
\begin{align}
	S_0 = - \sum_{n=1}^{N_{\text{b}}} p_n\log_{2} p_n,
\end{align}
where $p_n$ is the fraction of points lying in cell $n$, and $N_{\rm b}$ is the total number of cells in the $(\theta, \varphi)$ grid ($N_{\rm b}=30^2$ in this case). Note that for the statistics to be reliable there must be a sufficiently large number of points compared to the number of cells, namely $N/N_b \gg 1$; in our case $N/N_b = 100$.

The Shannon entropy strongly depends on the chosen grid, and its maximum value is $\log_2(N_{\rm b})$, corresponding to a uniform distribution over the $(\theta,\varphi)$ grid. 
Thus, let us change the normalization such that the entropy lies in the interval $[0,1]$,
\begin{align}
	S = - \frac{1}{\log_{2}(N_{\text{b}})}\sum_{n=1}^{N_{\text{b}}} p_n\log_{2} p_n.
\end{align}

\begin{figure}
	\centering

		\includegraphics[width=0.85\linewidth]{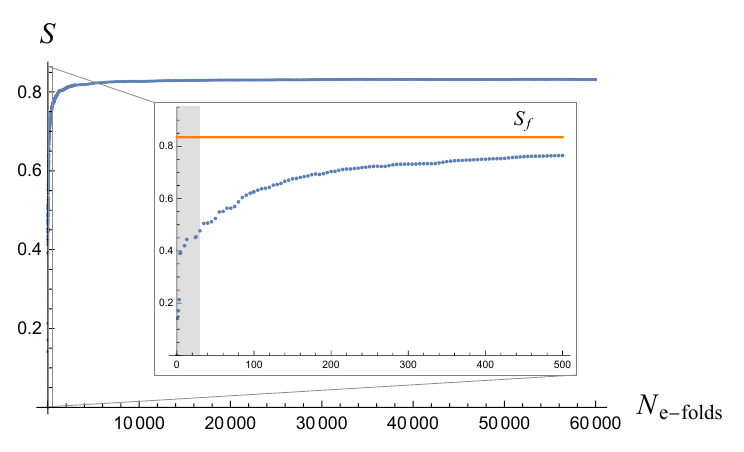}
		\caption{Numerical evolution of the Shannon entropy as a function of the number of $e$-folds, with a zoomed-in inset up to 500 $e$-folds. The horizontal orange line shows the saturation value, while the shaded gray region indicates the range of validity of the BKL regime,
        and its boundary with the white region corresponds to $N_{e\text{-folds}}=30$, when the system reaches the Planck scale.}
		\label{fig:shannon_evolution}

\end{figure}

The evolution of $S$ for an initially sharply peaked distribution is shown in Fig.~\ref{fig:shannon_evolution}.
Initially, since the distribution is localized, $S$ is very close to zero, but the dynamics causes $S$ to rapidly grow and then asymptote to a maximal value $S \approx 0.83$.
The evolution of the entropy is very well captured (especially at times after a few hundred $e$-folds of contraction) by
a power-law in time of the form
\begin{align}\label{fitting_curve_Shannon}
	S(t)=S_f+\frac{S_f- S_i}{\left(1+ A\, t \right)^{\gamma}},
\end{align}
where $S_f \approx 0.83$ represents the final (saturation) value of the entropy,
while $S_i$ is the initial value (close to $0$). However, we note that the best fit for the parameters $\gamma$ and $A$
is not universal, and depends on the initial state.

The maximal value for $S$ is precisely the (normalized) Shannon entropy for the final invariant distribution (shown in Fig.~\ref{fig:late_evolution}). As expected, since all initial distributions asymptote to the same final distribution, at late times $S$ will always tend to 
$S \to {S_f}\approx  0.83$.
This is also true for different choices of the initial cell in the $(\theta,\varphi)$ plane, and for different initial values of $P$ in the
commented range $P(t_0)\in[40,100]$. It is interesting to note that the asymptotic value is smaller than 1 (the value of $S$ for the uniform distribution), which is due to the presence of some clustering in the final distribution at certain angles $(\theta,\varphi)$.

However, within the 30 $e$-folds that it takes to reach the Planck scale and during which
classical general relativity can be trusted, $S$ does not reach by far its asymptotic value $S_f \approx 0.83$. As seen in Fig.~\ref{fig:shannon_evolution},
within this interval of time, the entropy (and hence the mixing) remains well below saturation.
Due to the initially rapid increase in $S$, the entropy reaches approximately 53\% of the asymptotic value by $t=30$,
that is, slightly more than half. Still, many $e$-folds beyond the first 30 are required to approach $S_f$,
since the rate of increase for $S$ rapidly decreases.
For example, 75\% of the asymptotic value $S_f$ is only reached after 100 $e$-folds, more than three times the 30 $e$-folds required to reach $\sim50\%$. Even in the most optimistic reading of this analysis, the Shannon entropy of a localized distribution can only get half-way
to the asymptotic value within the first 30 $e$-folds before reaching the Planck scale, and this finite-evolution time does not allow the chaotic properties to fully develop.

\subsection{Kullback-Leibler divergence}\label{sec_KLdivergence}

\begin{figure}
		\centering
		\includegraphics[width=0.85\textwidth]
        {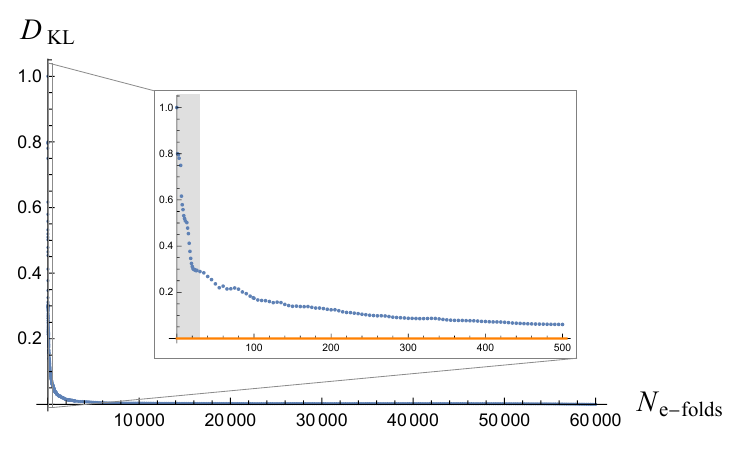}
		\caption{Numerical evolution of the Kullback-Leibler evolution as a function of the number of $e$-folds, with a zoomed-in inset up to 500 $e$-folds. The horizontal orange line shows the saturation value 0, while the shaded gray region indicates the range of validity of the BKL regime and its boundary with the white region corresponds to $N_{e\text{-folds}}=30$, when the system reaches the Planck scale.}
		\label{fig:KL_divergence}
\end{figure}

The fact that the chaotic features of the dynamics are not fully realized before reaching the Planck scale after 30 $e$-folds of contraction is also demonstrated by calculating the Kullback–Leibler divergence, which measures how far two phase-space distributions are from each other.
In this case we use the Kullback–Leibler divergence to determine the distance between the distribution for the state at a given time, and the final invariant distribution (which is shown in Fig.~\ref{fig:late_evolution}, and described in more detail in Sec.~\ref{sec.invariant}).
This is a useful complement to the Shannon entropy because, while $S$ in some way measures the distance of a distribution to the uniform distribution (which maximizes $S$), the Kullback–Leibler divergence directly quantifies the distance to the final invariant distribution reached by the dynamics, which is more relevant in this context than the uniform distribution.

The Kullback–Leibler divergence measuring the distance to the final invariant state is defined as
\begin{align}\label{def_KL}
	D_{\rm KL-0}=\sum_{n=1}^{N_b} p_n\log\left(
	\frac{p_n}{q_n}
	\right),
\end{align}
where $p_n$ and $q_n$ are the fraction of points lying in the $n$-th cell for the state under consideration and for the final invariant state, respectively. (In general, the sum is restricted to cells with nonzero $q_n$, but for the final invariant state in our grid all $q_n \neq 0$.)

The Kullback–Leibler divergence also depends on the grid that is used to partition the $(\theta,\varphi)$ plane.
As with the Shannon entropy, we normalize it dividing it
by its value for the initial sharply peaked state $D_{\rm KL-0}(t_0)$, which is the maximal value attained during the evolution,
\begin{align}\label{norm_KL}
	D_{\rm KL}=\frac{1}{D_{\rm KL-0}(t_0)}\sum_{n=1}^{N_b} p_n\log\left(
	\frac{p_n}{q_n}
	\right).
\end{align}

In this way $D_{\rm KL}$ takes values in $[0,1]$.
The evolution of $D_{\rm KL}$ is then shown in Fig.~\ref{fig:KL_divergence}. By construction, its initial value is 1, and it decreases towards zero as the system approaches the invariant state.  Within the regime of validity of the BKL behavior (i.e., during the first 30 $e$-folds of contraction during which the curvature remains below the Planck scale and general relativity can be trusted), the divergence decreases to $D_{\rm KL}\approx 0.29$, which still remains quite far from the final invariant state. (Note, in particular, that to decrease $D_{\rm KL}$ by another factor of 2 requires a total amount of over 140 $e$-folds of contraction.) 

Therefore, the Kullback–Leibler divergence shows similar features as the Shannon entropy: even though there is a reasonably rapid approach to the final invariant state, a distribution that is initially strongly localized still remains quite far from the final state after 30 $e$-folds of contraction, at which point the classical description of the BKL dynamics in terms of general relativity can no longer be trusted.

\subsection{Fourier analysis}
\label{sec_Fourier}

A Fourier analysis is a suitable method to quantify the level of structure in the distribution:
a relatively smooth spectrum indicates a well-mixed state, whereas peaks at specific Fourier modes indicate residual structure.
Focusing on the $(\theta,\varphi)$ section of the phase space, which topologically is
a two-dimensional torus $[0, 2\pi]\times[0, 2\pi]$, the discrete Fourier transform of the distribution reads
\begin{align}\label{def_Fourier_discrete}
	{\cal F}(k_1,k_2)=\frac{1}{N}\sum_{r=1}^N 
	e^{-i\left(k_1 \theta_r + k_2 \varphi_r\right)},
\end{align}
where $k_1$ and $k_2$ are the Fourier modes conjugate to $\theta$ and $\varphi$ respectively,
$N$ is the total number of points in the distribution,
while $(\theta_r,\varphi_r)$ are the coordinates of each point. Then, the power spectrum is simply
\begin{align}\label{def_power_spectrum}
	{\cal P}(k_1,k_2)=|{\cal F}(k_1,k_2)|^2.
\end{align}

\begin{figure}
	\centering
	
	\begin{subfigure}{0.48\textwidth}
		\centering
		\includegraphics[width=\textwidth]{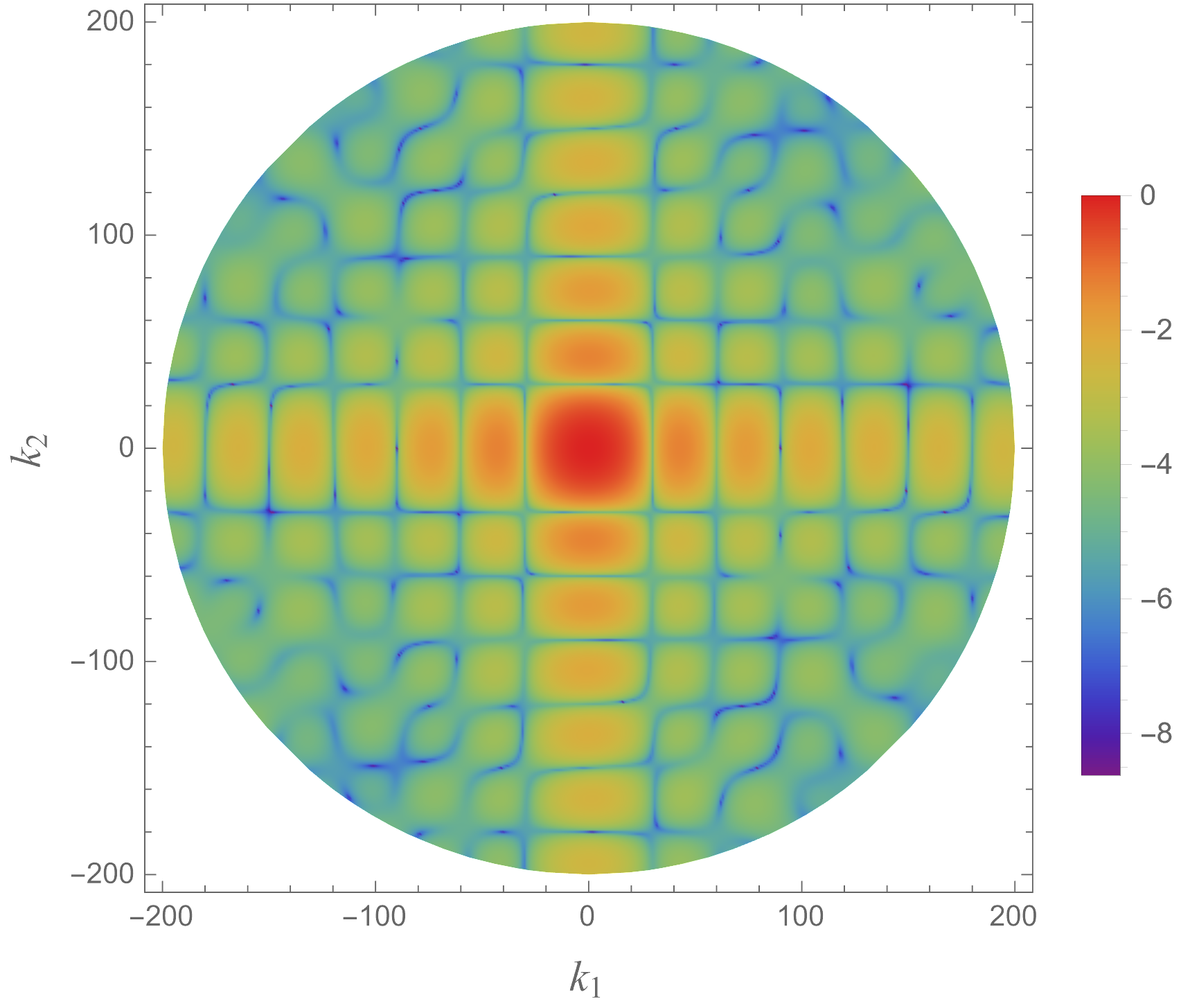}
		\caption{Initial state.}
		\label{fig:contour_initial}
	\end{subfigure}
	\hfill
	\begin{subfigure}{0.48\textwidth}
		\centering
		\includegraphics[width=\textwidth]{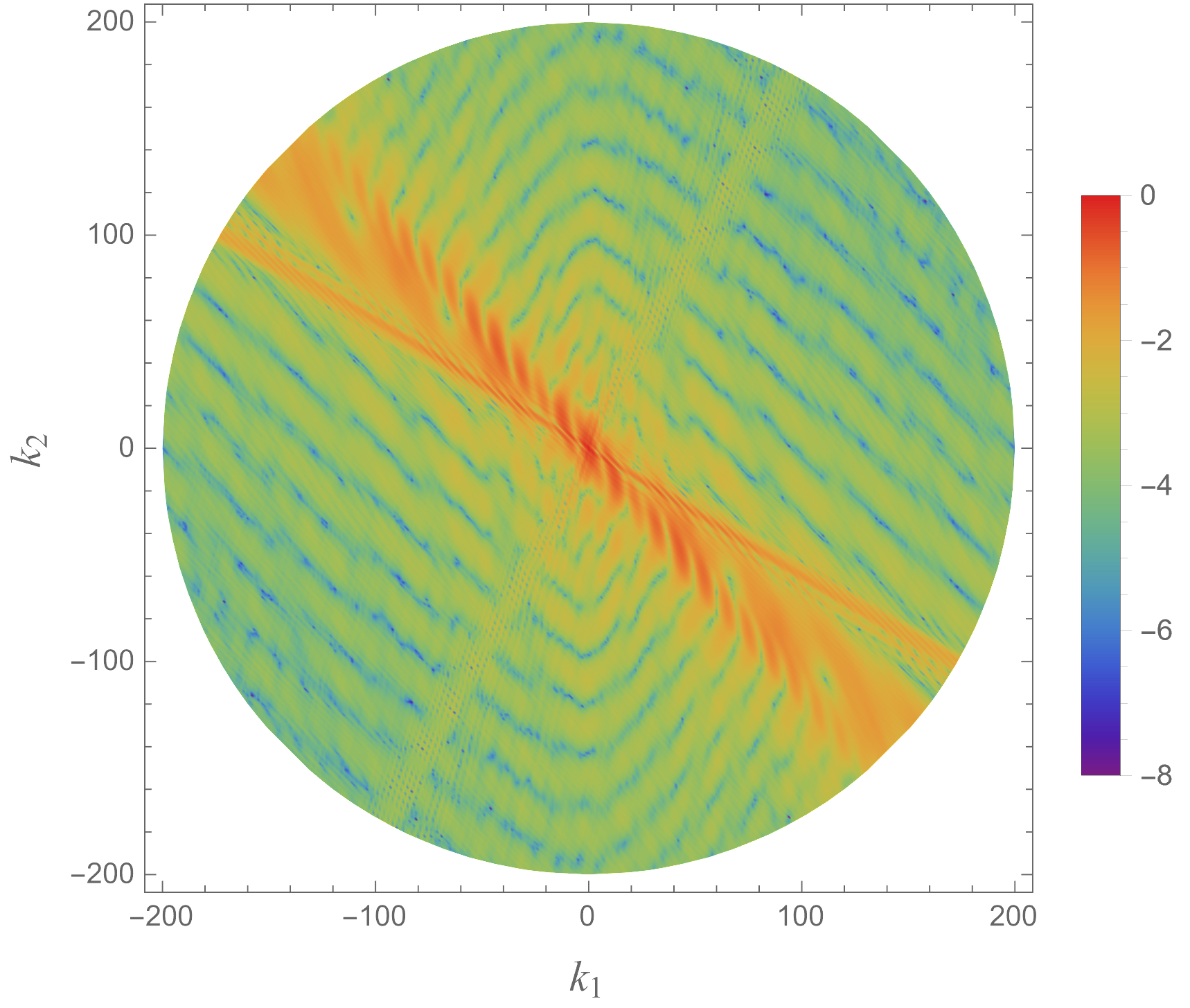}
		\caption{$30$ $e$-folds (Planck scale).}
		\label{fig:contour_limit}
	\end{subfigure}
	\vspace{0.5cm}
	
	\begin{subfigure}{0.85\textwidth}
		\centering
		\includegraphics[width=\textwidth]{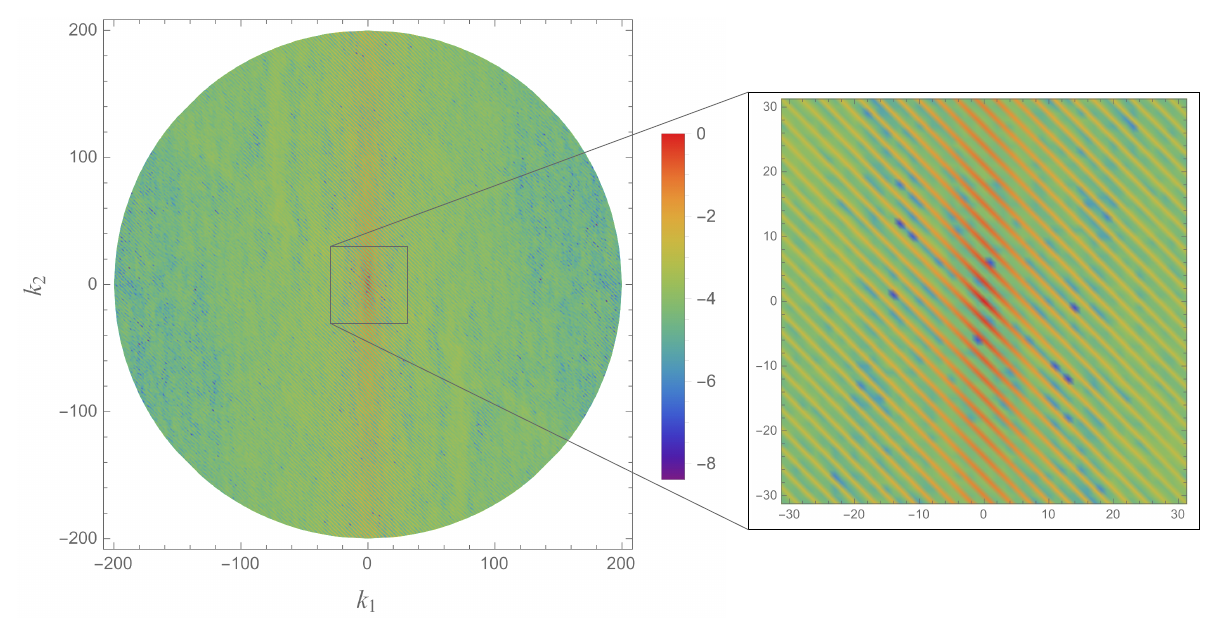}
		\caption{$60\, 000$ $e$-folds (final invariant state), with a zoomed-in inset up to $k_1,k_2\leq 30$.}
		\label{fig:contour_final}
	\end{subfigure}
	
	\caption{Contour plots in logarithmic scale of the Fourier power spectrum ${\cal P}(k_1,k_2)$ at different stages of the evolution.}
	\label{fig:contour_plot}
\end{figure}

Contour plots of the power spectrum in logarithmic scale are shown in Fig.~\ref{fig:contour_plot} for three different stages of the evolution.
The power spectrum for the initial distribution at $t=0$ is broad and highly structured, as can be seen in Fig.~\ref{fig:contour_initial},
and as expected for a strongly localized distribution. The dominant peak is centered at the origin $(k_1,k_2)=(0,0)$, demonstrating that the most important features are contained in the large-scale (low-frequency) modes, while the repeated minima and maxima are the
characteristic pattern of a distribution with a natural length scale. When the Planck scale is reached after $30$ $e$-folds of contraction,
most of the large-scale structure has disappeared, but there remains small-scale structure as can be seen in Fig.~\ref{fig:contour_limit}, so the spectrum is not yet fully smooth. Finally, Fig.~\ref{fig:contour_final} shows the power spectrum for the final invariant distribution. The spectrum has become smoother, suggesting a near scale-invariant spectrum, but it is not completely isotropic since some preferred directions remain. These preferred directions reflect the presence of clusters in the final distribution, as shown in Fig.~\ref{fig:late_evolution}. In fact, magnifying a subregion of Fig.~\ref{fig:contour_final} shows brighter diagonal features associated with the clusters around the angles $\theta,\varphi = 0, 2\pi/3, 4\pi/3, 2\pi$, as well as more pronounced vertical structures due to the clusters being narrower in the $\varphi$ direction. We will further discuss the features of this invariant state in Sec.~\ref{sec.invariant}.

\begin{figure}
	\centering
	\includegraphics[width=0.81\linewidth]{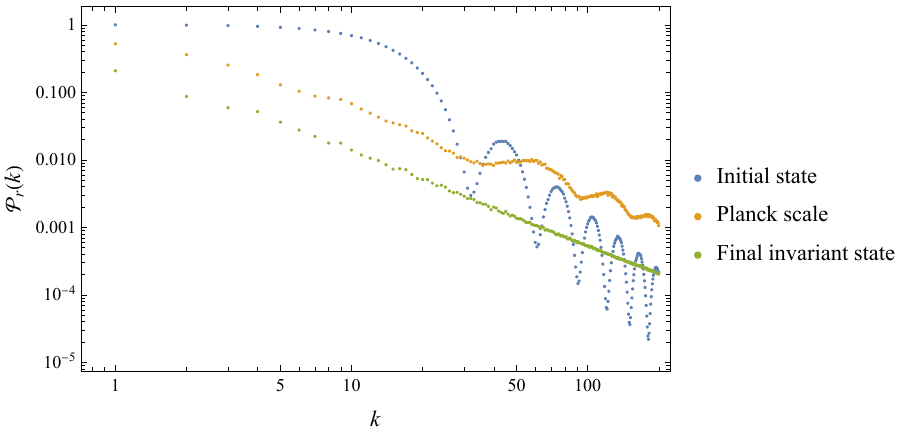}
	\caption{Radial Fourier power spectrum ${\cal P}_r(k)$ in log-log scale at three stages: initial state, Planck scale (end of validity of the BKL conjecture), and at the final invariant state.}
	\label{fig:radial_spectra}
\end{figure}

Another interesting quantity is the 
radial (isotropic) power spectrum ${\cal P}_r(k)$ obtained by averaging over points with the same radius in Fourier space,
that is,
\begin{equation}\label{def_radial}
	{\cal P}_r(k)=\langle {\cal P}(k_1,k_2)\rangle_{\sigma_k},
\end{equation}
where $\sigma_k$ is the collection of lattice points $(k_1,k_2)$ with $\sqrt{k_1^2 + k_2^2} = k$ (numerically, this is computed by defining $\sigma_k$ as the set of lattice points satisfying $k\leq \sqrt{k_1^2+k_2^2} \leq k+\epsilon$, and choosing $\epsilon > 0$
such that there are enough points in $\sigma_k$ for a meaningful average). This radial power spectrum is plotted in Fig.~\ref{fig:radial_spectra},
for the same three stages as above: for the initial state, at the Planck scale after 30 $e$-folds, and for the final invariant state that the distribution asymptotes to at late times. Initially, the power spectrum is strongly concentrated at small wave numbers, indicating that most of the information is contained in large-scale structures, with a rapid decay towards higher $k$,
reflecting the fact that the distribution is initially smooth and with very little fine-scale content.
After 30 $e$-folds, the power spectrum has decreased at small $k$ but increased at large $k$,
showing that the structure is now more strongly concentrated at smaller scales than initially.
Finally, for the invariant distribution there is structure at all scales and the power spectrum follows a straight line,
as will be discussed in more detail in Sec.~\ref{sec.invariant}.

\begin{figure}
		\centering
		\includegraphics[width=0.81\textwidth]{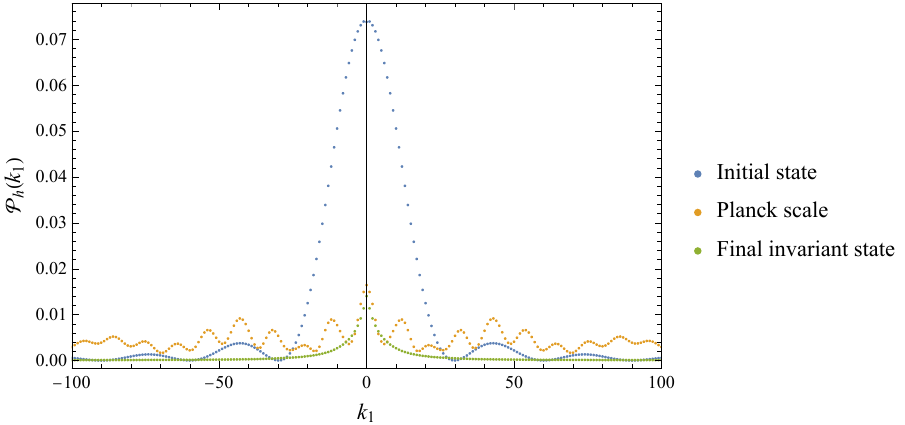}
		\caption{Horizontal power spectrum ${\cal P}_h(k_1)$  (defined as the averaged ${\cal P}(k_1,k_2)$ across each line $k_1=\text{const.}$) at the three stages.
        }
		\label{fig:horizontal_spectrum}	
\end{figure}

\begin{figure}
	\centering
	\includegraphics[width=0.81\textwidth]{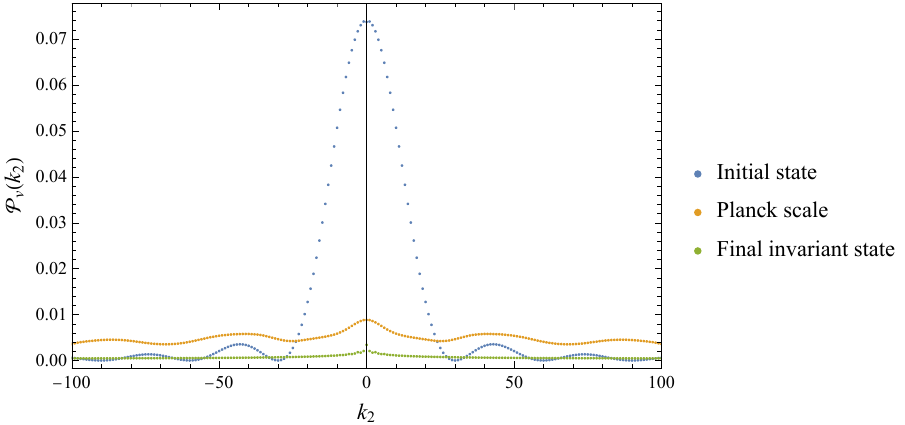}
	\caption{Vertical power spectrum ${\cal P}_v(k_2)$ (defined as the averaged ${\cal P}(k_1,k_2)$ across each line $k_2=\text{const.}$) at the three stages.}
	\label{fig:vertical_spectrum}
\end{figure}

Similarly, it is also possible to compute the horizontal and vertical power spectra, 
denoted by ${\cal P}_h(k_1)$ and ${\cal P}_v(k_2)$,
by taking the average of ${\cal P}(k_1,k_2)$ along each line $k_1=\text{const.}$ and $k_2=\text{const.}$, respectively. These are shown in Figs.~\ref{fig:horizontal_spectrum} and \ref{fig:vertical_spectrum}.

All this analysis clearly shows that after 30 $e$-folds (at the Planck scale), although the structure of the initial distribution in $(\theta,\varphi)$ has progressively been transferred from larger to smaller scales, there remains considerable structure at all scales: the orange curve in all figures lies above the green curve for the final invariant state at all scales. In short, at this time the system remains far from the final mixed state, and the loss of information about the initial configuration remains incomplete.

\vspace{0.4cm}

\vspace{0.4cm}

\section{Characterization of the invariant final distribution}
\label{sec.invariant}

In this section we describe in a little more detail the asymptotic invariant distribution of the system expressed in terms of the $(\theta,\varphi)$ variables, as it is shown in Fig.~\ref{fig:late_evolution}. While the existence of  such an invariant state is well known in the literature \cite{Barrow_Chernoff:1983, Cornish:1996hx, Cornish:1996yg, Imponente:2003hx}, to the best of our knowledge it has not yet been studied using variables where the two relevant chaotic directions are compactified.

It is important to emphasize that the same invariant state (up to numerical fluctuations) is obtained from any initial distribution, whether the initial distribution is sharply peaked or not. In Sec.~\ref{sec-quantitative}, we only considered initial conditions corresponding to a sharply peaked distribution, but all states eventually tend to the same final invariant state.
As an example of this, we show in App.~\ref{sec_randomstate} that an initially uniform distribution is also dynamically driven to the same final state. There are of course differences during the initial transient, since the trajectories are not identical at early $e$-folds, but, after this transient regime, all distributions eventually converge to the same invariant state.

The first thing to note about the invariant final distribution is that, as shown in Fig.~\ref{fig:late_evolution}, it is not a uniform distribution. Instead, the distribution is denser near the points $\theta,\varphi=0, 2\pi/3, 4\pi/3, 2\pi$, indicating the influence of an underlying repelling structure that organizes the dynamics. This is also very clear from the histogram of Fig.~\ref{fig:histogram}, which exhibits Gaussian-like peaks centered at these angular locations. Due to these clusters, the (normalized)
Shannon entropy of the invariant state does not attain its maximum possible value of $S=1$,
but rather $S_f \approx 0.83$. This is nonetheless a relatively high value close to 1, indicating a high degree of mixing.

A Fourier analysis also reflects the presence of these peaks in the final invariant distribution. For example, as can be seen in Fig.~\ref{fig:contour_final}, the power spectrum is not completely isotropic, as there remain some preferred directions due to the peaks. Further, a zoom of the same figure shows brighter diagonal features associated with these clusters, and also a more pronounced vertical structure due to the peaks being narrower in the $\varphi$ direction.

One way to isolate the effects due to these peaks is to consider an idealized scenario in which the distribution is replaced by Dirac delta functions centered at the locations of the peaks $\theta,\varphi=0, 2\pi/3, 4\pi/3, 2\pi$. In this case, the Fourier transform can be computed analytically, and the power spectrum is
\beq
{\cal P}(k_1,k_2)\propto \left[1 + 2 \cos \left(\frac{2\pi}{3} (k_1+k_2)\right)\right]^2.
\ee
This expression depends only on the combination $k_1+k_2$, which explains the appearance of diagonal lines satisfying $k_1+k_2=m$, for $m$ an integer, in Fig.~\ref{fig:contour_final}. Further, the spectrum is enhanced at multiples of three (whenever the integer $m$ is divisible by 3), 
which precisely correspond to the observed brighter lines. 
(It is also possible to improve on this idealization by replacing the Dirac delta functions by Gaussians with different widths in the $\theta$ and $\varphi$ directions, with similar though more complicated results.)

Another important feature of the final invariant state that is made clear by the Fourier analysis is that the radial power spectrum, defined in \eqref{def_radial} and shown in Fig.~\ref{fig:radial_spectra}, is always nonzero, showing that there is structure at all scales in the final state. Moreover, the power spectrum follows a straight line in the logarithmic plot, denoting a power-law scaling
\beq
{\cal P}_r(k)\propto k^{\delta}, 
\ee
where the best fit gives $\delta\approx -1.38$. This result 
is characteristic of fractals and self-similar structures.

\section{Discussion}
\label{sec-disc}

The BKL conjecture provides an asymptotic description of the approach to generic spacelike singularities in general relativity,
where the evolution of neighboring points decouple, and each point follows the chaotic Mixmaster equations of motion.
However, since general relativity does not have a natural length scale, there is no universal scale at which the BKL
scenario will universally hold. Rather, it occurs asymptotically, and the curvature scale of its onset is different for each spacetime. 
Since candidate extensions of general relativity usually introduce a finite ultraviolet scale,
most commonly linked to the Planck scale, beyond which the classical Einstein equations are expected to break down,
the following questions naturally arise:
how long can the BKL scenario last, and how much chaos can be generated by the Mixmaster dynamics at each point before 
classical general relativity ceases to be valid?

In order to provide an answer, we have first assumed that the BKL conjecture holds, so the evolution of each spatial point can be accurately described by
the Einstein equations corresponding to the Mixmaster (diagonal vacuum Bianchi~IX) model.
We then have argued that, for physically relevant astrophysical and cosmological spacetimes, there can be at most $\sim 30$ $e$-folds of contraction from the initial onset of the BKL behavior to the Planck scale. This bound is derived from rather conservative assumptions, and a more realistic analysis would give a smaller number of $e$-folds of contraction that would further limit the development
of the chaotic features.

Starting from a sharply peaked distribution on the phase space, we observe that it asymptotically evolves to the final invariant state that acts as an attractor for all distributions with some initial spread.
Further, the development of chaotic behavior over the finite interval of 30 $e$-folds of contraction can be quantified in terms of
the Shannon entropy, the Kullback-Leibler divergence, 
and a Fourier spectral analysis.
We found that these quantities all evolve monotonically towards their asymptotic values, but nonetheless they all consistently show that,
after only 30 $e$-folds of evolution,
the distribution remains far from the final invariant state.
Indeed, at the Planck scale, (i) the initially localized distribution spreads only partially over the reduced phase space
(see Fig.~\ref{fig:planck_limit})
and clearly remains far from the final invariant distribution (shown in Fig.~\ref{fig:late_evolution}),
(ii) basin boundaries retain a smooth (non-fractal) structure (Fig.~\ref{fig:basins2}),
(iii) the Shannon entropy has only reached about half of its asymptotic value (Fig.~\ref{fig:shannon_evolution}),
(iv) the normalized Kullback-Leibler divergence remains significantly different from zero (Fig.~\ref{fig:KL_divergence}),
and (v) the Fourier spectrum (Fig.~\ref{fig:contour_limit}) preserves substantial large-scale coherence,
instead of exhibiting the fully developed scale-invariant behavior characteristic of the invariant repeller.
While this discussion concerns a phase-space distribution that is initially sharply peaked, similar results hold more generally, including the case of an initial uniform distribution that is presented in App.~\ref{sec_randomstate}.

These results suggest that the chaotic properties commonly associated with the BKL conjecture and with the Mixmaster dynamics should be regarded primarily as asymptotic mathematical properties of the Einstein equations, rather than as phenomena that will necessarily arise within the regime of validity of general relativity. As a result, the classical evolution appears insufficient to erase a substantial fraction of the information encoded in the initial state---whether this information is ultimately preserved or destroyed crucially depends on the physics governing the high-curvature regime beyond the classical cutoff.

This observation has potentially important implications for a broad class of quantum-gravity and modified-gravity scenarios. For example, there exist many theories that predict a nonsingular bounce connecting a pre-bounce contracting phase to a post-bounce expanding phase, and these may 
preserve considerably more information about the pre-bounce geometry than previously expected, provided that the bouncing phase itself does not introduce additional sources of chaotic mixing. 
In particular, in effective loop quantum cosmology the cosmic bounce in the Mixmaster universe occurs over a short timescale \cite{Wilson-Ewing:2010lkm, Corichi:2015ala, Giovannetti:2019ewe} and can be described as a different type of Kasner transition \cite{Chiou:2007mg, Wilson-Ewing:2017vju, Wilson-Ewing:2018lyx} (this transition rule has since been found to hold for nonsingular cosmic bounces arising in many theories \cite{deCesare:2019suk, LeFloch:2020pdf}). Given the small number of Kasner transitions on either side of the bounce, and although analysis of more general inhomogeneous models, such as Gowdy spacetimes \cite{Garay:2010sk, Brizuela:2009nk, Brizuela:2011ps}, point towards an enhancement of inhomogeneities through the bounce, the chaotic features of the Mixmaster dynamics can be expected to have a limited impact on the amplitude of the inhomogeneities of the original spacetime, as also argued in \cite{Wilson-Ewing:2017vju, Wilson-Ewing:2018lyx, Blackmore:2023wiv}. Alternatively, if the curvature remains bounded due to quantum gravity, then the potential walls may become ineffective at inducing further Kasner transitions, which
would limit the chaotic features of the Mixmaster dynamics \cite{Bojowald:2004ra, Antonini:2018gdd}.
Likewise, in gravitational collapse, the amount of mixing of classical information occurring before quantum-gravity effects become relevant may be substantially smaller than what the asymptotic BKL picture alone would suggest.

Finally, the methodology introduced here provides a quantitative framework for studying finite-time chaotic mixing in gravitational dynamics. Although the present analysis has been restricted to the vacuum Bianchi IX model, the methods used here to quantify the emergence of chaos can be naturally and directly extended to include matter fields---these likely will not yet be completely negligible at finite curvature \cite{Ali:2017qwa, Brizuela:2024ggl, Muzammil:2025nuv}---as well as to modified gravity theories, including those with effective quantum-gravity corrections, such as the models described in Refs.~\cite{Brizuela:2022uun, Bojowald:2023fas, Bojowald:2023sjw}.

\section*{Acknowledgments}
D.~B. and S.~F.~U. acknowledge financial support from the Basque Government Grant \mbox{IT1977-26}, and by the Grant PID2021-123226NB-I00 (funded by
MCIN/AEI/10.13039/501100011033 and by ``ERDF A way of making Europe'').
E.~W.-E.~was supported in part by the Natural Sciences and Engineering Research Council of Canada.

\appendix
\newpage
\section{Initial conditions}
\label{sec:app.initial}

In this appendix, we explain how the initial conditions for the simulations have been chosen.
To start, we review the definition of the Misner variables in Sec.~\ref{sec:app.Misner}, and explain how the compact angular variables are defined in terms of the Misner variables in Sec.~\ref{sec:app.angles}. Then, in Sec.~\ref{sec:app.fix} we describe the steps we follow to fix the initial conditions.

\subsection{Misner variables}
\label{sec:app.Misner}

The Misner variables $(\alpha,\beta_+,\beta_-)$ are defined by the following
parameterization of the directional scale factors \cite{Misner:1969ae, Misner:1969hg}:
\begin{align}
\label{aiintermsofmisner}
a_1 =e^{\alpha}\,e^{\beta_++\sqrt{3}\beta_-}, \qquad 
a_2 =e^{\alpha}\,e^{\beta_+-\sqrt{3}\beta_-} , \qquad 
a_3 =e^{\alpha}\,e^{-2\beta_+},
\end{align}
with $a=e^\alpha$ being the mean scale factor. Inverting these relations gives
\begin{align}
	\label{def_Misner_var}
	\alpha:=\frac{1}{3}\ln\left(a_1 a_2 a_3\right),\qquad
	\beta_+:=\frac{1}{6}\ln\left(
	\frac{a_1 a_2}{a_3^2}	\right),\qquad
	\beta_-:=\frac{1}{2\sqrt{3}}\ln\left(
	\frac{a_1}{a_2}
	\right).
\end{align}
Then, their (rescaled) dimensionless conjugate momenta read
\footnote{From the Einstein-Hilbert action, it can be verified that the conjugate momenta to the Misner variables are
$\tilde p_\alpha= \pi^2 r_0^2 p_\alpha/\kappa$,
$\tilde p_+= \pi^2 r_0^2 p_+/\kappa$, and
$\tilde p_-= \pi^2 r_0^2 p_-/\kappa$,
where $\kappa=8\pi G$ and $G$ is Newton's constant.
}
\begin{align}
	\label{def_momenta_Misner_var}
		&
		p_\alpha
        =-12r_0 \frac{a^2\dot{a}}{N},
		\qquad
		p_+=\frac{2r_0a^3}{N}\left(
		\frac{\dot{a}_1}{a_1}+
		\frac{\dot{a}_2}{a_2}-
		\frac{2\dot{a}_3}{a_3}
		\right),
		\qquad
		p_-=\frac{2\sqrt{3}r_0a^3}{N}\left(
		\frac{\dot{a}_1}{a_1}-
		\frac{\dot{a}_2}{a_2}
		\right),
\end{align}
where, at this stage, the dot stands for a derivative with respect to any time variable $t$.

In terms of the Misner variables, the constraint \eqref{constraint} simplifies to
\begin{align}\label{constraint_misner}
	p_+^2+p_-^2-p_\alpha^2+e^{4\alpha}
	U(\beta_+,\beta_-)=0,
\end{align}
where 
\begin{align}\label{def_potential}
U(\beta_+,\beta_-):=48\left(
e^{-8\beta_+}
-4e^{-2\beta_+}\cosh 2\sqrt{3}\,\beta_-
+4e^{4\beta_+}\sinh^2 2\sqrt{3}\,\beta_-
	\right).
\end{align}
As mentioned previously, the terms in the constraint that are quadratic in momenta can be interpreted
as kinetic terms, while the term $e^{4\alpha} U(\beta_+,\beta_-)$ plays the role of a potential.

From this point on, as in the main body of the paper, we will work in the gauge
$t=-\alpha$ (equivalently $t=-\ln a$), such that the singularity ($a\to 0$) is located at $t=+\infty$.
This choice corresponds to the lapse
\begin{align}
	\label{def_lapse}
N=\frac{12e^{3\alpha}r_0}{p_\alpha}.
\end{align}
In order to completely fix the gauge, $p_\alpha$
is obtained by solving the
constraint \eqref{constraint_misner},
\begin{align}\label{def_p_alpha}
	p_\alpha=\sqrt{p_+^2+p_-^2+e^{-4t}
	U(\beta_+,\beta_-)},
\end{align}
where the positive root has been chosen so the universe is collapsing, that is, $\dot{a}<0$.

Therefore, to solve the system in this gauge, there only remains to 
find the solutions for
the variables $\beta_\pm(t)$ and $p_\pm(t)$, which follow the equations
\begin{align}\label{eqm_beta_p}
\begin{aligned}
 &\frac{d\beta_+}{dt}=
 \frac{p_+}{p_\alpha},
 \qquad \qquad
 &\frac{d p_+}{dt}= 
-\frac{e^{-4t}}{2p_\alpha}
 \frac{\partial U}{\partial \beta_+},
  \\[5pt]
 &\frac{d\beta_-}{dt}=
 \frac{p_-}{p_\alpha},
 \qquad \qquad
 &\frac{d p_-}{dt}= -\frac{e^{-4t}}{2p_\alpha}
 \frac{\partial U}{\partial \beta_-},
 \end{aligned}
\end{align}
with $U$ and $p_\alpha$ as given in \eqref{def_potential} and \eqref{def_p_alpha}, respectively.
Before providing the initial conditions for this set of equations, let us comment on some relevant features.

First, we note that the fiducial length $r_0$ is completely absent from the equations \eqref{eqm_beta_p}, and also from \eqref{def_p_alpha}.
Therefore, the evolution of the 
Misner variables is completely independent of $r_0$.
In particular, the solutions $(\beta_{\pm}(t),p_{\pm}(t))$ contain no information
about $r_0$. Thus, $r_0$ should be regarded simply as a fiducial length that needs
to be specified only when reconstructing the spacetime geometry, but that does not affect the dynamics.

Second, the equations of motion \eqref{eqm_beta_p} are invariant under the simultaneous transformation
\begin{equation}\label{trans1}
t\to t-\ln c,
\end{equation}
or equivalently $a\to c a$, and
\begin{equation}\label{trans2}
p_{\pm}\to c^2 p_\pm.
\end{equation}
(Note that, from \eqref{def_p_alpha}, this
implies $p_\alpha \to c^2 p_\alpha$.)
Therefore, given a solution $(\beta_{\pm}(t),p_{\pm}(t))$,
one can define a one-parameter family of solutions 
\beq \label{one-parameter}
\beta_{\pm}^{(c)}(t)=\beta_{\pm}(t+\ln c),\quad p^{(c)}_{\pm}(t)=c^2 p_{\pm}(t+\ln c),
\ee
for any constant $c>0$, that are related by this transformation.
By explicitly reconstructing the metric \eqref{def_metric}
using \eqref{aiintermsofmisner} and \eqref{def_lapse}--\eqref{def_p_alpha},
it is possible to see that, if one 
also rescales $r_0 \to r_0/c$, then
all these solutions yield the same spacetime geometry. This comes from the fact that
both the combination $a r_0$ and the lapse $N$ \eqref{def_lapse} are invariant
under the transformation \eqref{trans1}--\eqref{trans2} combined with the rescaling $r_0\to r_0/c$.
Note that it is possible to rescale $r_0$ since, as mentioned above, $r_0$
is not fixed by the Misner variables.
Hence, up to a rescaling of the fiducial length, all the solutions of the family
\eqref{one-parameter}
define exactly the same spacetime geometry, and thus are physically equivalent.
As a result, when providing initial conditions for the system \eqref{eqm_beta_p},
we will take into account that the physically meaningful information about the geometry is encoded in 
quantities that are invariant under the transformation \eqref{trans1}--\eqref{trans2}.

\subsection{Angular variables}
\label{sec:app.angles}

Next, we rewrite the four dynamical variables $\beta_\pm$ and $p_\pm$ in a polar form,
in terms of two angular variables $\theta$ and $\varphi$, and two positive norm variables $\beta$ and $P$,
as follows,
\begin{align}\label{angular_variables_Misner}
\beta_+=\beta\cos{\varphi},\qquad \beta_-=\beta\sin{\varphi},\qquad 
p_+=P\cos\theta,\qquad p_-=P\sin\theta,
\end{align}
where $\beta\in [0,\infty)$ is the modulus of the vector $(\beta_+,\beta_-)$, and $\varphi\in[0, 2\pi]$ is its angle with respect to the $\beta_+$-axis, while $P\in[0,\infty)$ and $\theta\in[0,2\pi]$ play the analogous role for the momenta $p_\pm$.
In terms of these polar variables, from \eqref{def_p_alpha}, $p_\alpha$ reads
\begin{align}\label{def_p_alpha_angular}
	p_\alpha=\sqrt{P^2+e^{-4 t}
	U(\beta_+,\beta_-)}.
\end{align}

Finally, it is worth noting that these angles are a compactified version of the $(u,v)$ parameters introduced in Ref.~\cite{Cornish:1996hx}:
\beq
u=-\frac{1}{2}\left(1+\sqrt{3}\cot\frac{\theta}{2}\right),\qquad
v=\frac{1+\sqrt{3}\cot\varphi}{1-\sqrt{3}\cot \frac{\theta}{2}}.
\ee
The chaotic nature of the Bianchi~IX model was proven by showing that the space of initial conditions in the $(u,v)$ plane has a fractal structure~\cite{Cornish:1996hx}; the same will be true for the space of initial conditions in the $(\theta,\varphi)$ plane.

\subsection{Fixing the initial conditions}
\label{sec:app.fix}

The initial conditions for a sharply peaked distribution in phase space are fixed in the following manner.
After the gauge-fixing of the time variable, there remain
four variables $(\beta,P, \theta,\varphi)$, but, since
the evolution equations \eqref{eqm_beta_p} explicitly depend on $t$, the choice of
the initial value $t=t_0$ is also dynamically relevant.
Note that, in the gauge under consideration
$t=-\ln a$, fixing $t_0$ is equivalent to fixing $a(t_0)=e^{-t_0}$.
Therefore, there are five independent initial data to be chosen:
$(t_0,\beta(t_0),P(t_0), \theta(t_0),\varphi(t_0))$.

In addition, 
it is convenient to use
variables that are invariant under the symmetry transformation \eqref{trans1}--\eqref{trans2} of the equations. 
Interestingly, three of the variables under consideration $(\theta,\varphi,\beta)$ are invariant, while
the transformation implies $a\to c a$ and $P\to c^2 P$. Therefore, $a$ and $P$ are not invariant, but their ratio
$P/a^2$ is.

In order to impose initial conditions for a sharply peaked distribution in phase space,
first the $(\theta,\phi)$ space $[0,2\pi]\times[0,2\pi]$ is split into a $30\times 30$ grid, and one cell in this grid is randomly selected. We then randomly select $N=300^2$
values of $\theta$ and $\varphi$ within that cell. This fixes the initial value of the angles, $\theta(t_0)$ and $\varphi(t_0)$.
Next, as is common in the literature, we will require that each solution in the distribution is initially in a Kasner regime, where the dynamics is
very well approximated by the vacuum Bianchi~I model. Specifically, to ensure this condition, we impose that $\beta(t_0)$ is
chosen such that the potential term in the constraint \eqref{constraint_misner} exactly vanishes:
\beq
U(\beta\cos\varphi,\beta\sin\varphi)\big|_{t=t_0}
=0.
\ee
In this way, since $U$ initially vanishes, the definition of $p_\alpha$ \eqref{def_p_alpha_angular}
simply implies that
\begin{equation}\label{pt0Pt0}
p_\alpha(t_0)=P(t_0).
\end{equation}

Once $\beta(t_0)$ is fixed,
it only remains to fix the initial values for $P(t_0)$ and $a(t_0)$.
To do this, it is helpful to briefly recall the physical evolution of the Bianchi~IX spacetime,
which has a recollapse at large volumes and then asymptotically tends to the singularity, where the BKL behavior emerges.
In the approach to the singularity, the duration (in terms of the time variable $t=-\ln a$) of the Kasner epochs becomes longer and longer (see Fig.~\ref{fig:evolution_scale_factors}, where each portion of the dynamics with
all three directional scale factors evolving monotonically is a Kasner epoch).
Thus, it is important to avoid starting too close to the singularity because, in that case,
there would be very few Kasner transitions, and this would bias the analysis
since the amount of mixing and the development of the chaotic features will grow with the number of transitions.
On the other hand, it is equally important to avoid starting too close to the recollapse, because in that regime the evolution is not approximated
by a sequence of Kasner epochs, and it is not relevant to describe the BKL behavior.

In conclusion, to describe a realistic scenario, at the initial time the system can be neither too close to the singularity, nor too close to the recollapse. To quantify this condition, we introduce the ratio
\beq
\rho=\frac{p_\alpha}{a^2}=\sqrt{e^{4 t} P^2+U},
\ee
which has several interesting features. 

First, from \eqref{pt0Pt0}, at the initial time $\rho$
takes takes the simple form
\begin{equation}
 \rho(t_0)=\frac{P(t_0)}{a^2(t_0)}.
\end{equation}
 
Second, $\rho$ is invariant under the symmetry transformation \eqref{trans1}--\eqref{trans2}.
Therefore, the initial value of $\rho$
is the dynamical variable that defines the spacetime geometry,
while, as long as $r_0$ is not fixed, the initial values of $a(t_0)$ and $P(t_0)$ 
by themselves do not have an invariant meaning,
and simply select a representative member of the family of solutions \eqref{one-parameter}.
That is, both the initial data sets $\{a(t_0)=e^{-t_0}, P(t_0)=P_0\}$
with fiducial length $r_0=R_0$,
and $\{a(t_0-\ln c)=c e^{-t_0}, P(t_0-\ln c)=c^2 P_0\}$ with fiducial length $r_0=R_0/c$ 
lead to the same spacetime geometry. The evolution of the corresponding Misner variables will
be related by the
symmetry transformation  \eqref{trans1}--\eqref{trans2}.

Third, we note that, from the recollapse to the singularity,
$\rho$ is monotonically increasing, as can be seen from the equations of motion \eqref{eqm_beta_p}:
\beq
\frac{d\rho}{dt}=2e^{2t}\,\frac{p_+^2+p_-^2}{p_\alpha}=2e^{2t}\frac{P^2}{p_\alpha}> 0.
\ee
Recall from \eqref{def_p_alpha} that $p_\alpha$ is positive during contraction. At the recollapse, $p_\alpha=0$
with finite $a$, so $\rho=0$; and it can be verified that, approaching the singularity, $a$ vanishes more rapidly
than $p_\alpha$, so $\rho \to \infty$ at the singularity. As a result, from the recollapse $\rho$ increases monotonically from 0 and diverges at the singularity.

In summary, to ensure that the initial conditions are set neither too close to the recollapse nor too close to the singularity, it is necessary to select an intermediate value for $\rho(t_0)$ as an initial condition, 
while the choice of $a(t_0)$, or equivalently of $P(t_0)$,
is not physically important, since different choices will give equivalent dynamics
simply related by \eqref{trans1}--\eqref{trans2}.
Following a common choice in the literature, it is convenient to set
$a_1(t_0)=1$, which, from \eqref{def_Misner_var}, implies
\begin{align}\label{ini_time}
t_0 = \beta\left(\sqrt{3}\sin\varphi + \cos\varphi\right),
\end{align}
and provides typical values for $a(t_0)$ of the order $a(t_0)\sim 0.1$.
Then, based on an extensive study of different numerical simulations, a reasonable range to maximize the number of Kasner transitions, yet remain far from the recollapse, is approximately $\rho(t_0) \in [4~000, 10~000]$, which, given the choice for $a(t_0)$, implies
$P(t_0)\in [40,100]$. We choose $P(t_0)$ randomly within this interval, taking the same value for $P(t_0)$ for all of the points lying in the phase-space distribution. It turns out that the qualitative results are essentially identical for simulations performed with different values of $P(t_0)$ lying within this range.

\section{The Kretschmann Scalar}
\label{sec.Kretschmann}

In terms of the Misner variables, the Kretschmann scalar $K = R_{abcd} R^{abcd}$ for the Bianchi~IX spacetime is:
\begin{align}
r_0^4 K &=\frac{256}{3} e^{-4\alpha}
\bigg[
e^{-16\beta_+}
+ e^{-4\beta_+}
- 2e^{-10\beta_+}\cosh\!\left(2\sqrt{3}\,\beta_-\right)
+2e^{8\beta_+}\left(
\cosh\!\left(8\sqrt{3}\,\beta_-\right)-\cosh\!\left(4\sqrt{3}\,\beta_-\right)
\right)
\nonumber \\[5pt]
&- 8e^{2\beta_+}
\cosh\!\left(2\sqrt{3}\,\beta_-\right)\sinh^2\!\left(2\sqrt{3}\,\beta_-\right)
\bigg]
+\frac{1}{216}
e^{-12\alpha}
\left(
\left(p_+^2+p_-^2\right)^2
+
p_+\,p_\alpha\left(p_+^2-3p_-^2\right)
\right)
\nonumber \\[5pt]
&
+\frac{8}{9}
e^{-8\alpha+4\beta_+}\,
p_+\left(p_+-p_\alpha\right)
+\frac{4}{9}
e^{-8\alpha+4\sqrt{3}\,\beta_-+4\beta_+}
\left(
p_+\left(p_\alpha-p_+ \right)+\sqrt{3}\,p_-\left( p_\alpha-5p_+\right)-6p_-^2\right)
\nonumber \\[5pt]
&
+\frac{4}{9}
e^{-8\alpha-4\sqrt{3}\,\beta_-+4\beta_+}
\left(
p_+\left(p_\alpha-p_+ \right)-\sqrt{3}\,p_-\left( p_\alpha-5p_+\right)-6p_-^2\right)
+\frac{2}{9}
e^{-8\alpha-8\beta_+}
\left(
3p_-^2
-
p_+\left(17p_+ + 4p_\alpha\right)
\right)
\nonumber \\[5pt]
&
+\frac{2}{9}
e^{-8\alpha+2\sqrt{3}\,\beta_- -2\beta_+}
\left(
3p_-^2
-2\sqrt{3}\,p_-\left(p_+ + p_\alpha\right)
+p_+\left(p_+ + 2p_\alpha\right)
\right)
\nonumber \\[5pt]
&
+\frac{2}{9}
e^{-8\alpha-2\sqrt{3}\,\beta_- -2\beta_+}
\left(
3p_-^2
+2\sqrt{3}\,p_-\left(p_+ + p_\alpha\right)
+p_+\left(p_+ + 2p_\alpha\right)
\right).
\end{align}

In the approach to the singularity corresponding to the limit $a=e^{\alpha}\to 0$,
the system is mostly in a Kasner epoch. At these times, the exponential terms containing the shape parameters $\beta_\pm$ are subleading,
and then $K$ can be approximated as
\beq
K\approx
\frac{e^{-12\alpha}}{216 \, r_0^4}
\left(
\left(p_+^2+p_-^2\right)^2
+
p_+\,p_\alpha\left(p_+^2-3p_-^2\right)
\right).
\ee
This expression can be rewritten in terms of the compactified variables,
\beq
K\approx
\frac{e^{-12\alpha}P^4}{108 \, r_0^4}
\sin^2\left(
\frac{3\theta}{2}
\right).
\ee
Note that these asymptotic expressions are only valid
for $\theta\neq 2\pi m/3$, with $m=0,1,2$.
(For the particular cases $\theta= 2\pi m/3$, with integer $m$, the leading-order term vanishes and the remaining dominant term goes as $e^{-8\alpha}$.)
Therefore, in general, the scaling of the Kretschmann curvature scalar in the approach to the singularity is $K\propto (a^3\,r_0 )^{-4}$, as stated in Sec.~\ref{sec-range}. 
Note also that this ignores the dependence on $P$, which is a reasonable approximation because (as can be verified numerically) $a$ varies much more rapidly than $P$ and, to leading order, $P$ is nearly constant.

\section{Alternative: Initial uniform distribution on phase space}
\label{sec_randomstate}

\begin{figure}
	\centering
	
	\begin{minipage}{0.48\textwidth}
		\centering
		\includegraphics[width=\textwidth]{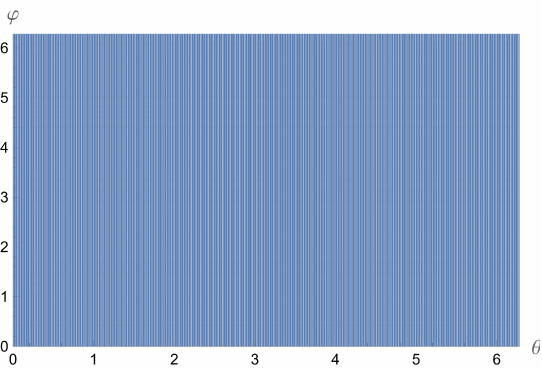}
		\caption{Initial uniform distribution in $(\theta,\varphi)$ space.}
		\label{fig:initial_distribution_reverse}
	\end{minipage}
	\hfill
	\begin{minipage}{0.48\textwidth}
		\centering
		\includegraphics[width=\textwidth]{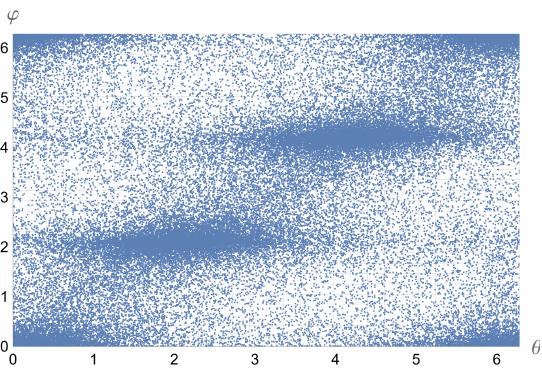}
		\caption{Final invariant distribution in $(\theta,\varphi)$ space starting from the initial uniform distribution.}
		\label{fig:final_distribution_reverse}
	\end{minipage}

\end{figure}

Instead of starting from a sharply peaked distribution, it is also interesting to take an initial distribution that randomly samples the entire $(\theta,\varphi)$ space
in a uniform way,
and let that distribution, shown in Fig.~\ref{fig:initial_distribution_reverse}, evolve under the Mixmaster dynamics. In this case, for each $(\theta(t_0),\varphi(t_0))$, the initial conditions for the remaining variables are chosen in the same way as for the sharply peaked state, as explained in App.~\ref{sec:app.fix}.
The result is that the distribution clusters around the same regions of $\theta,\varphi=0,2\pi/3,4\pi/3,2\pi$. These clusters become progressively more refined, and the distribution eventually converges to the same invariant final state, as can be seen by comparing Fig.~\ref{fig:final_distribution_reverse} and Fig.~\ref{fig:late_evolution}.

\begin{figure}
	\centering
	\includegraphics[width=0.85\linewidth]{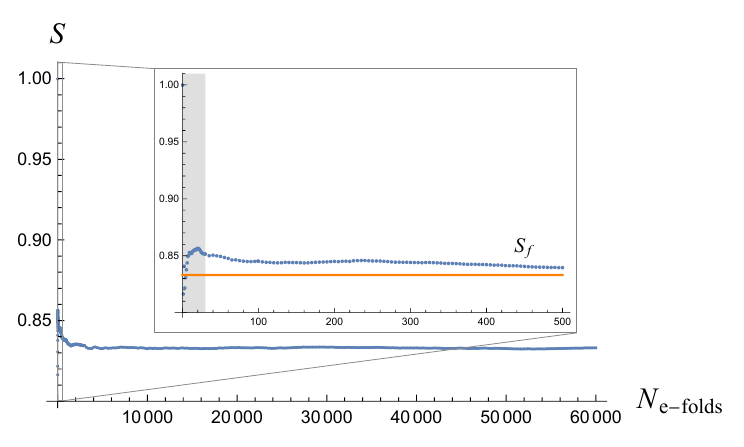}
	\caption{Numerical evolution of the Shannon entropy as a function of the number of $e$-folds for an initially uniform distribution covering the full phase space, with a zoomed-in inset up to 500 $e$-folds. The horizontal orange line shows the saturation value, while the shaded gray region indicates the range of validity of the BKL regime and its boundary with the white region corresponds to $N_{e\text{-folds}}=30$, when the system reaches the Planck scale.}
	\label{fig:shannon_reverse}
\end{figure}

This convergence is also supported by evaluating the Shannon entropy, as shown in Fig.~\ref{fig:shannon_reverse}. Since the initial state is fully random, the Shannon entropy starts at the maximum possible value of $1$, and then decreases to lower values and eventually approaches a plateau (although not monotonically, contrary to distributions that are initially sharply peaked), where it asymptotes to the Shannon entropy of the final invariant state, $S_f \approx 0.83$.

\begin{figure}
    \centering
    \includegraphics[width=0.85\linewidth]{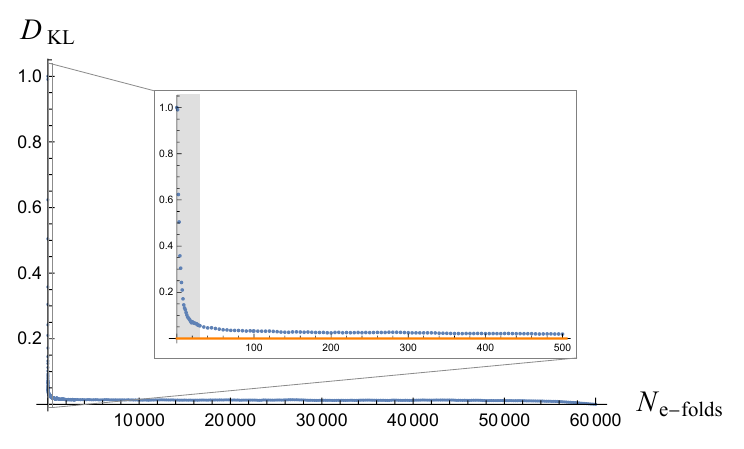}
    \caption{Numerical evolution of the Kullback-Leibler evolution as a function of the number of $e$-folds for an initially uniform distribution, with a zoomed-in inset up to 500 $e$-folds. The horizontal orange line shows the saturation value 0, while the shaded gray region indicates the range of validity of the BKL regime and its boundary with the white region corresponds to $N_{e\text{-folds}}=30$, when the system reaches the Planck scale.}
    \label{fig:KL_full_space}
\end{figure}

The Kullback–Leibler divergence \eqref{norm_KL}, shown in Fig.~\ref{fig:KL_full_space}, also demonstrates that the initial uniform state is rapidly driven towards the final invariant state: $D_{\rm KL}$ decays considerably faster for this uniform initial distribution the initially random state, compared to a distribution that is initially sharply peaked, and quite rapidly asymptotes to zero.

Finally, the power spectrum also rapidly tends to the power spectrum of the final invariant state, and, after sufficient mixing, it is indistinguishable from the power spectrum shown in Fig.~\ref{fig:contour_final}.


\raggedright

\end{document}